\documentclass[11pt]{article}
\usepackage[margin=1in]{geometry}
\usepackage{amsmath,amssymb,amsthm,mathtools,bm,booktabs,multirow}
\usepackage{algorithm,algpseudocode}
\usepackage{graphicx,xcolor}
\usepackage{tikz}
\usetikzlibrary{arrows.meta}
\usepackage[colorlinks=true,linkcolor=blue!55!black,citecolor=blue!55!black,urlcolor=blue!55!black]{hyperref}
\usepackage[round,authoryear]{natbib}
\usepackage{setspace}
\usepackage{placeins}
\newcommand{\Sref}[1]{\ref{#1}}
\newcommand{\Mref}[1]{\ref{#1}}

\newif\ifblind
\blindfalse   

\newtheorem{theorem}{Theorem}
\newtheorem{proposition}{Proposition}
\newtheorem{corollary}{Corollary}
\newtheorem{lemma}{Lemma}
\theoremstyle{definition}
\newtheorem{assumption}{Assumption}

\theoremstyle{remark}
\newtheorem{remark}{Remark}

\newcommand{\Le}{L_\varepsilon}
\newcommand{\cL}{\mathcal L}
\newcommand{\cA}{\mathcal A}
\newcommand{\GV}{\mathcal G}
\newcommand{\Mor}{\mathcal M}

\newcommand{\diag}{\operatorname{diag}}
\newcommand{\blkdiag}{\operatorname{blkdiag}}
\newcommand{\pa}{\operatorname{pa}}
\newcommand{\E}{\mathbb E}
\newcommand{\NOt}{NO$_2$}
\newcommand{\PMt}{PM$_{2.5}$}

\title{Dynamic bigraphical Mat\'ern--Whittle processes: learning dependence graphs among spatial-temporal fields}
\ifblind
\author{}
\else
\author{Alokesh Manna\thanks{Department of Statistics, Texas A\&M University, College Station, Texas 77843, U.S.A.} \and Debangan Dey\footnotemark[1]}
\fi
\date{}

\begin{document}
\maketitle

\begin{abstract}
Many studies record several variables at the same sites over a sequence of periods, such as pollutant concentrations at monitoring stations or case counts of infectious diseases by district. A central question concerns the dependence between the variables and its change over time. We develop dynamic bigraphical Mat\'ern--Whittle (DBMW) processes for such data. Each variable has a spatial field that evolves by autoregression. A graph over the variables couples the temporal innovations of the fields. Its links appear and vanish over time, and a link may change sign. A graph Laplacian on the sites gives the spatial structure, and a canonical exponential family links the fields to Gaussian, count or binomial responses. The joint distribution of the latent fields over all sites, variables and periods factorizes over the eigenmodes of the graph Laplacian, and its normalizing constant is free of the graph and of the dynamics. Given the fields, the posterior of the time-varying signed graph is therefore exact, by forward--backward recursions. The estimator has three steps: a fit without a graph, which also gives the switching rates with an interval, the exact graph posterior, and a joint refit. In simulations, DBMW recovers the graph with higher average precision than graphical lasso, time-varying graphical lasso and Poisson log-normal networks, and it predicts withheld data better than a standard spatio-temporal model that fits each variable separately, with a Mat\'ern field and autoregressive evolution, by integrated nested Laplace approximation. In weekly air-quality data from 288 U.S. sites, it recovers the seasonal change of sign between ozone and nitrogen dioxide. In German surveillance data on seven diseases, the district-level links stay positive and stable through the COVID-19 pandemic.
\end{abstract}

\noindent\textbf{Keywords:} Conditional independence, Dynamic graphical model, Exponential family, Gaussian Markov random field, Graph Laplacian, Hidden Markov model, Laplace approximation.

\section{Introduction}\label{sec:intro}

Many scientific questions concern how several quantities measured over space depend on one another, and whether this dependence changes over time. Air-quality networks record several pollutants at hundreds of sites each week \citep{epa_aqs}. Disease surveillance reports case counts of several notifiable conditions by district and period \citep{rki_survstat}. Ecological surveys count many species at the same sites each season, brain-imaging studies record signals in many regions over time, and climate data sets give several variables on a grid each month. In each case an observation carries a variable, a location and a time. The scientific interest lies mainly in the relations between the variables, after accounting for each variable's own spatial and temporal structure. A dependence graph summarizes such relations. Its nodes are the variables, and a link joins two variables that remain dependent given all the others. A second question concerns the stability of this graph over the observation period.

\begin{figure}[!htbp]
\centering
\includegraphics[width=\linewidth]{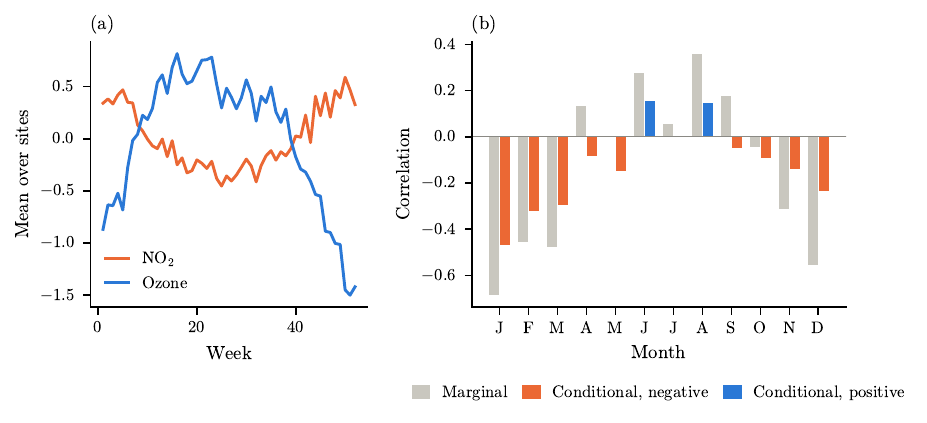}
\caption{Ozone and \NOt\ at U.S. monitoring sites in 2023. (a) Weekly mean over the 288 sites of the standardized log concentrations. (b) Correlation across sites of the week-to-week changes of the two pollutants, averaged over the weeks of each month, at the 54 sites with observations of all five variables. Grey: marginal correlation. Coloured: partial correlation given temperature, CO and \PMt.}\label{fig:intro}
\end{figure}

Figure~\ref{fig:intro} gives an example with ozone and nitrogen dioxide (\NOt) at 288 U.S. monitoring sites in 2023. The two pollutants follow opposite seasonal cycles (Figure~\ref{fig:intro}a). Across sites, the correlation of their week-to-week changes is negative from November to March and positive from June to September (Figure~\ref{fig:intro}b). Conditional on temperature, carbon monoxide (CO) and fine particles (\PMt), the winter association stays negative, and the summer association stays positive in June and August, at a smaller size. Atmospheric chemistry explains this pattern \citep{seinfeld2016}. In winter, nitric oxide emitted with \NOt\ destroys ozone. In summer, sunlight drives the production of ozone from nitrogen oxides. One dependence graph for the whole year averages the two regimes into a single weak link. A model for such data therefore needs links that appear, vanish and change sign over time.

Questions of this kind call for a joint model with four properties. Each variable needs its own spatial field with its own temporal evolution. The dependence between variables needs an interpretable graph, with links free to change over time, in sign as well as in presence. The responses may be measurements, counts or binomial records. Estimation must remain feasible with hundreds of thousands of latent values. Existing approaches meet these requirements only in part. Multivariate spatial models from cross-covariance functions \citep{gneiting2010, genton2015} or multivariate conditional autoregressions \citep{gelfand2003, jin2005} make every pair of variables dependent and rarely scale beyond a few variables. Dynamic multivariate spatial models let the spatial processes evolve over time through state-space dynamics \citep{gelfand2005dynamic}, and basis-function models reduce the dimension of multivariate spatio-temporal areal data \citep{bradley2015}. Neither gives a graph over the variables. Spatio-temporal models from stochastic partial differential equations (SPDEs) \citep{lindgren2011, cameletti2013}, fitted by integrated nested Laplace approximation \citep{rue2009}, handle one or a few fields without such a graph. Graph estimation, on the other hand, mostly concerns vectors without spatial structure. Examples are the graphical lasso \citep{friedman2008}, its time-varying versions \citep{zhou2010, kolar2010, hallac2017}, Poisson log-normal networks for counts \citep{chiquet2019} and Bayesian dynamic graphical models \citep{carvalho2007}. Among the Bayesian models, hidden Markov models switch between network states in brain imaging \citep{warnick2018}, and change-point models detect a change of graph through the COVID-19 pandemic \citep{franzolini2024}. Time-varying graphs also appear in space, for a graph over brain locations whose structure changes over time \citep{greenewald2017}, but there the graph joins sites, not variables. Graph-based multivariate spatial processes encode conditional independence between whole processes. The graph is known in graphical Gaussian processes \citep{dey2022graphical} and in graph-constrained models for multivariate functional data \citep{dey2025graphconstrained}. It is learned from data by the basis graphical lasso \citep{krock2023} and in the bigraphical Mat\'ern--Whittle (BMW) process \citep{dey2026bmw}. In the BMW process, a graph Laplacian on the sites gives each variable a Mat\'ern-type field, and a directed acyclic graph couples the fields. These graph-based processes are static and Gaussian. No current method combines spatial fields, temporal dynamics, a graph with switching signs and non-Gaussian responses.

\begin{figure}[!htbp]
\centering
\definecolor{posc}{HTML}{2A78D6}\definecolor{negc}{HTML}{EB6834}
\begin{tikzpicture}[x=1cm,y=1cm,
  tile/.style={inner sep=0pt,draw=black!40,line width=0.3pt},
  vnode/.style={circle,draw=black!70,line width=0.5pt,minimum size=5mm,inner sep=0pt,font=\small},
  posl/.style={-{Latex[length=2mm]},line width=0.9pt,draw=posc},
  negl/.style={-{Latex[length=2mm]},line width=0.9pt,draw=negc},
  off/.style={line width=0.6pt,draw=black!30,dash pattern=on 1.5pt off 1.5pt},
  ar/.style={-{Latex[length=2mm]},line width=0.6pt,draw=black!55}]
\def\ts{1.45}
\foreach \t/\x in {1/2.6,2/5.9,3/9.2,4/12.5}{
  \node at (\x,7.05) {$t=\t$};
  \foreach \i/\y in {1/6.0,2/4.3,3/2.6}{
    \node[tile] (w\i\t) at (\x,\y) {\includegraphics[width=\ts cm]{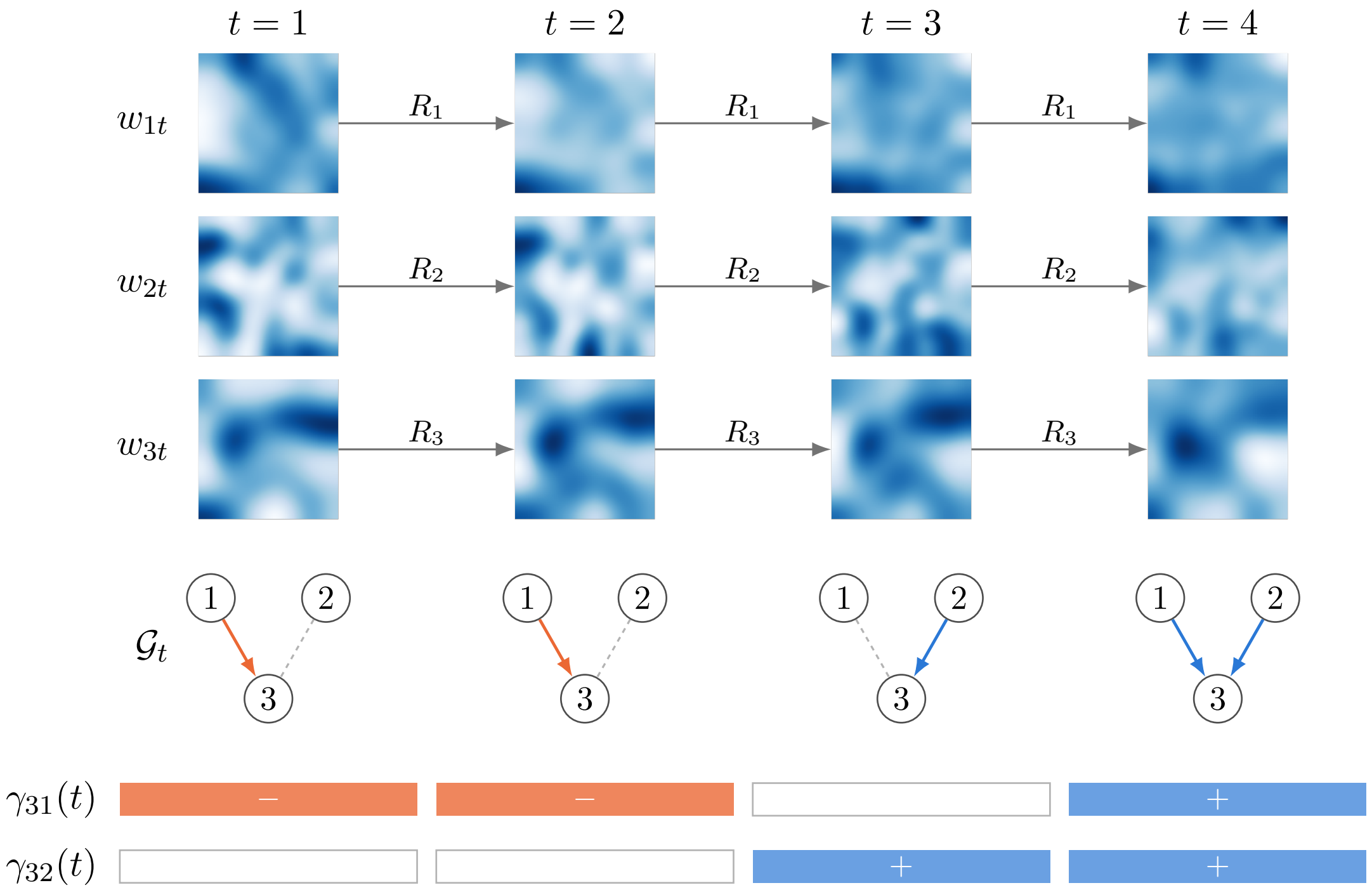}};
  }
}
\foreach \i/\y in {1/6.0,2/4.3,3/2.6}{ \node[anchor=east] at (1.7,\y) {$w_{\i t}$}; }
\foreach \a/\b in {1/2,2/3,3/4}{
  \foreach \i in {1,2,3}{ \draw[ar] (w\i\a.east) -- node[above,font=\footnotesize,inner sep=1pt]{$R_\i$} (w\i\b.west); }
}
\foreach \t/\x/\ea/\eb in {1/2.6/negl/off,2/5.9/negl/off,3/9.2/off/posl,4/12.5/posl/posl}{
  \node[vnode] (a\t) at (\x-0.6,1.05) {1};
  \node[vnode] (b\t) at (\x+0.6,1.05) {2};
  \node[vnode] (c\t) at (\x,0.0) {3};
  \draw[\ea] (a\t) -- (c\t);
  \draw[\eb] (b\t) -- (c\t);
}
\node[anchor=east] at (1.7,0.55) {$\mathcal G_t$};
\foreach \row/\y/\lab/\pat in {1/-1.05/{\gamma_{31}(t)}/{-1,-1,0,1},2/-1.75/{\gamma_{32}(t)}/{0,0,1,1}}{
  \node[anchor=east] at (0.95,\y) {$\lab$};
  \foreach \v [count=\k] in \pat {
    \pgfmathsetmacro{\xx}{2.6+3.3*(\k-1)}
    \ifnum\v=1 \fill[posc!70] (\xx-1.55,\y-0.17) rectangle (\xx+1.55,\y+0.17); \node[font=\footnotesize,text=white] at (\xx,\y) {$+$};
    \else\ifnum\v=-1 \fill[negc!80] (\xx-1.55,\y-0.17) rectangle (\xx+1.55,\y+0.17); \node[font=\footnotesize,text=white] at (\xx,\y) {$-$};
    \else \draw[black!30,line width=0.5pt] (\xx-1.55,\y-0.17) rectangle (\xx+1.55,\y+0.17); \fi\fi
  }
}
\end{tikzpicture}
\caption{Conceptual overview of the DBMW process. Each variable $i$ has a spatial field $w_{it}$ in every period $t$, and each field evolves by autoregression ($R_i$) from its previous state. The graph $\mathcal G_t$ couples the temporal innovations of the fields. Its links switch between negative (orange), absent (dashed) and positive (blue) association. The bars show the link states $\gamma_{31}(t)$ and $\gamma_{32}(t)$.}\label{fig:concept}
\end{figure}

This paper develops such a framework, the dynamic BMW (DBMW) process. Its central construction separates time from dependence. Each variable's latent field evolves by autoregression in time. The innovations of the fields, the part of each field that its past leaves unexplained, form a multivariate spatial process at every period. A graph over the variables couples these innovations. Each link of the graph follows a Markov chain over three states: negative, absent and positive association. The rates at which links appear and disappear, the switching rates, are estimated from the data. A canonical exponential family links the fields to the responses. The spatial structure of the innovations follows the BMW process, and this choice makes exact computation possible. A comparison of two graphs requires the normalizing constant of the latent fields, the factor that makes their density integrate to one. Under the BMW structure this constant has a closed form that depends neither on the graph nor on the dynamics, because the variables are coupled in a directed, triangular way and the dynamics enter through a transformation with unit determinant. All spatial operators are also functions of one graph Laplacian on the sites, so the whole process splits into independent pieces, one per eigenvector of this Laplacian. Other spatial processes could be coupled in the same directed way, but without a shared eigenbasis they lose this splitting, and the computation becomes far more costly. With the BMW structure, the posterior over time-varying signed graphs is exact given the fields. Figure~\ref{fig:concept} gives a conceptual overview. In the figure, the link between variables 3 and 1 is negative in the first two periods, absent in the third and positive in the fourth. A static graph would join variable 3 to both variables at all times, with one weak coefficient each.

The paper makes three contributions. First, it gives a general framework of signed dynamic graphs over spatial fields, for any canonical exponential family. The fit without a graph, on/off switching, the static graph and the static Gaussian BMW process are special cases (\S\ref{sec:model}). Second, it derives the structure of the process. For $p$ variables, $n$ sites and $T$ periods, the joint distribution of the latent fields factorizes over the eigenvectors of the graph Laplacian into independent $p$-variate state-space models. The log-determinant of the $npT$-dimensional precision matrix equals $T$ times the static closed form, for any graph sequence and any dynamics. Absent links give exact conditional independences, localized in time (\S\ref{sec:structure}). Third, it develops an estimator in three steps. A fit without a graph gives the switching rates with a calibrated interval. Forward--backward recursions then give the exact posterior of the signed graph, one hidden Markov model per variable. A joint refit of fields, graph and parameters completes the estimator. The first two steps form a faster version, DBMW-X (\S\ref{sec:estimators}, Figure~\ref{fig:pipeline}).
Section~\ref{sec:model} defines the model, and Section~\ref{sec:structure} gives its structure. Section~\ref{sec:inference} develops the estimator. Section~\ref{sec:sim} reports the simulations, and Section~\ref{sec:applications} the two applications. The Supplementary Material contains the proofs, further details of the estimator, eight further simulation scenarios and further results of the applications.

\section{Dynamic bigraphical Mat\'ern--Whittle processes}\label{sec:model}

\subsection{Data and notation}
The data involve a set of locations $\mathcal S=\{s_1,\dots,s_n\}$, a set of variables $\{1,\dots,p\}$ and a set of time points $\{1,\dots,T\}$. The latent field of variable $i$ at time $t$ is $w_{it}\in\mathbb R^n$, with entries $w_{it}(s)$. We write $w_t=(w_{1t}^\top,\dots,w_{pt}^\top)^\top\in\mathbb R^{pn}$ and $w_{1:T}=(w_1^\top,\dots,w_T^\top)^\top$. The response $Y_{sit}$ records variable $i$ at location $s$ and time $t$. Entries may be missing.

\subsection{The spatial building block}\label{sec:bmw}
Following \citet{dey2026bmw}, $\Le$ denotes the density-normalized graph Laplacian of a nearest-neighbour graph on $\mathcal S$, scaled to unit mean diagonal. It is symmetric and positive semidefinite, with eigendecomposition $\Le=U\Lambda U^\top$, $\Lambda=\diag(\lambda_1,\dots,\lambda_n)$ and $U=(u_1,\dots,u_n)$. We call the eigenvectors $u_h$ the spatial eigenmodes. A small eigenvalue $\lambda_h$ belongs to a smooth eigenmode, and a large one to a rough eigenmode. The coefficients $u_h^\top w$ of a field $w$ are its spectral coordinates. For variable $i$, with range parameter $\kappa_i>0$ and smoothness parameter $\alpha_i>0$, write
\[
d_i(\lambda)=(\kappa_i^2+\lambda)^{\alpha_i/2},\qquad \cL_{ii}=d_i(\Le)=(\kappa_i^2I_n+\Le)^{\alpha_i/2}.
\]
Given a directed acyclic graph $\GV$ on $\{1,\dots,p\}$, whose links $i\leftarrow j$ respect a fixed ordering ($j<i$), and coupling coefficients $b_{ij}$, the static BMW field solves
\begin{equation}\label{eq:bmw}
(\kappa_i^2I_n+\Le)^{\alpha_i/2}\,w_i+\sum_{j\in\pa(i)}b_{ij}\,w_j=z_i,\qquad z_i\sim N(0,\tau_i^2I_n),\qquad i=1,\dots,p,
\end{equation}
independently over $i$. A variable $j$ with a link $i\leftarrow j$ is a parent of $i$, and $i$ is a child of $j$. In matrix form $\cL w=z$, where $\cL$ is block lower triangular with diagonal blocks $\cL_{ii}$ and off-diagonal blocks $b_{ij}I_n$, and $z\sim N(0,D_\tau)$ with $D_\tau=\diag(\tau_1^2,\dots,\tau_p^2)\otimes I_n$. Then $w\sim N(0,Q^{-1})$ with $Q=\cL^\top D_\tau^{-1}\cL$. The parameter $\nu_i=\alpha_i-d/2$ governs the smoothness of variable $i$ in dimension $d$ \citep[Proposition~1]{dey2026bmw}. For each eigenmode $h$ we use the $p\times p$ matrices
\begin{equation}\label{eq:Lh}
L_h=\diag\{d_1(\lambda_h),\dots,d_p(\lambda_h)\}+B,\qquad \tilde L_h=D_{\tau,0}^{-1/2}L_h,\qquad Q_h=\tilde L_h^\top\tilde L_h,
\end{equation}
where $B=(b_{ij})$ is strictly lower triangular and $D_{\tau,0}=\diag(\tau_i^2)$.

\subsection{Temporal dynamics on the innovations}\label{sec:dynamics}
The spatio-temporal process applies \eqref{eq:bmw} not to the field itself but to its temporal innovation. For $t=1,\dots,T$, with $w_{i0}=0$,
\begin{equation}\label{eq:stbmw}
(\kappa_i^2I_n+\Le)^{\alpha_i/2}\,\big(w_{it}-R_i\,w_{i,t-1}\big)+\sum_{j\in\pa_t(i)}b_{ij}(t)\,\big(w_{jt}-R_j\,w_{j,t-1}\big)=z_{it},\qquad z_{it}\overset{\text{iid}}{\sim}N(0,\tau_i^2I_n).
\end{equation}
Here $R_i$ is the temporal transition of variable $i$, and $\pa_t(i)$ and $b_{ij}(t)$ are the parents and coefficients of $i$ in the graph $\GV_t$ at time $t$. With $\eta_t=w_t-Rw_{t-1}$ and $R=\blkdiag(R_1,\dots,R_p)$, \eqref{eq:stbmw} states that
\[
\cL_t\,\eta_t=z_t,\qquad \eta_t\sim N(0,Q_t^{-1}),\qquad Q_t=\cL_t^\top D_\tau^{-1}\cL_t,
\]
independently over $t$. Here $\cL_t$ is the BMW operator of \S\ref{sec:bmw} with graph $\GV_t$ and coefficients $B_t=\{b_{ij}(t)\}$. At every time point, the innovation of the multivariate field is a static BMW field. The fields themselves carry memory through $R$.

\begin{assumption}[spectral transition]\label{ass:R}
Each $R_i=r_i(\Le)$ is a function of the spatial operator.
\end{assumption}

We use scalar persistence, $r_i(\lambda)\equiv\phi_i\in(-1,1)$, so each variable has its own autoregressive coefficient $\phi_i$, its persistence, as in the spatio-temporal model of \citet{cameletti2013}. Spectral persistence, $r_i(\lambda)=\{1+\Delta\,d_i(\lambda)\}^{-1}$, is the implicit Euler step of the evolution equation $(\partial_t+\cL_{ii})w_i=\dot z_i$, under which rough spatial components decay faster than smooth ones. With $T=1$ the model is the static BMW process. With $p=1$ it is an autoregressive Mat\'ern-type field in the style of \citet{cameletti2013}, with the SPDE mesh replaced by a graph Laplacian.

\subsection{Signed switching graph}\label{sec:graphprior}
The candidate links $(i,j)$, with $j<i$, form a set $\mathcal C$. By \eqref{eq:stbmw}, the innovation of variable $i$ moves with $-b_{ij}(t)$ times that of variable $j$. The state $\gamma_{ij}(t)\in\{-1,0,+1\}$ of link $(i,j)$ at time $t$ records a negative, absent or positive association, that is, the sign of $-b_{ij}(t)$. Each sign has its own strength,
\begin{equation}\label{eq:edgecoef}
b_{ij}(t)=\begin{cases}-\beta^+_{ij}+\delta_{ij}(t), & \gamma_{ij}(t)=+1,\\ 0, & \gamma_{ij}(t)=0,\\ \beta^-_{ij}+\delta_{ij}(t), & \gamma_{ij}(t)=-1,\end{cases}\qquad \beta^\pm_{ij}\sim N^+(0,s^2),\qquad \delta_{ij}(t)\overset{\text{iid}}{\sim}N(0,s_\delta^2),
\end{equation}
where $N^+$ denotes the half-normal law and $s_\delta\ll s$. The periods in which a link has one sign share one strength. This pooling matters, because a single time point carries little information about a coefficient.

The states of each link follow a Markov chain, independently over links. With sparsity parameter $\pi$ and switching rates $\epsilon_{\rm on}$ and $\epsilon_{\rm off}$,
\begin{equation}\label{eq:edgechain}
\begin{gathered}
\Pr\{\gamma_{ij}(1)=\pm1\}=\pi/2,\qquad \Pr(\pm1\to0)=\epsilon_{\rm off},\\
\Pr(0\to\pm1)=\begin{cases}\epsilon_{\rm first}/2 & \text{before the first presence},\\ \epsilon_{\rm on}/2 & \text{after it},\end{cases}
\end{gathered}
\end{equation}
with $\epsilon_{\rm first}=\epsilon_{\rm off}\,\pi/(1-\pi)$, and all other transitions keep the state. A change of sign therefore passes through absence. The chain separates a candidate that has never been a link from a link that has been present and is now absent. Most candidates are never links, and their entry is governed by the sparsity $\pi$. The rates $\epsilon_{\rm on}$ and $\epsilon_{\rm off}$ describe how real links switch: the duration of a present link and the time to its return. They are estimated from the data (\S\ref{sec:rates}). For computation, each state is augmented by the set of signs used so far, which gives eight states per link (Supplement~\Sref{sec:chain}). Section~\ref{sec:graphstep} explains the role of this memory.

Together, \eqref{eq:edgecoef} and \eqref{eq:edgechain} form a dynamic spike-and-slab prior. In each period a coefficient is either exactly zero, the spike, or drawn from a slab of either sign, and the Markov chain carries the spike or slab state from one period to the next, as in dynamic variable selection \citep{rockova2021}. Spike-and-slab priors also select covariates and autoregressive lags jointly in time-series regression \citep{manna2025lags}.

The construction contains three special cases. With one admissible sign per link, the model has on/off switching. With $\epsilon_{\rm off}=0$, every link keeps its first state, and the graph is static. A static link with a random sign and a half-normal strength has a $N(0,s^2)$ coefficient, so the static case is the spike-and-slab prior \citep{mitchell1988, george1993} of \citet{dey2026bmw}. We call the estimator of this static case DBMW-0.

\begin{remark}[Interpretation of the sign]\label{rem:sign}
If $j$ is a parent of $i$ at time $t$ and the two variables have no common child at $t$, the $(i,j)$ entry of $Q_{h,t}$ is $\tau_i^{-2}d_i(\lambda_h)b_{ij}(t)$, and the partial correlation of the innovations of $i$ and $j$ in every eigenmode has the sign of $-b_{ij}(t)$, the sign $\gamma_{ij}(t)$. A positive link thus means that the two innovations move together, given all other innovations at that time.
\end{remark}

\begin{remark}[Shared switching rates]\label{rem:switch}
One pair of rates, and one sparsity $\pi$, serve all candidate links. Each link contributes only $T-1$ transitions, typically with one or two switches, so an unpooled rate would sit near the boundary of its parameter space. Link-specific, covariate-dependent and time-varying rates change only the factors of the transition matrix in Proposition~\ref{prop:fb}, and the forward--backward computation stays exact at the same cost (Supplement~\Sref{sec:chain}). Section~\ref{sec:sim} and Supplement~\Sref{sec:other} examine the common rate with link-specific rates in the truth.
\end{remark}

\subsection{Observation model}\label{sec:obs}
Given the latent fields, the responses are conditionally independent with a canonical-link exponential-family distribution and linear predictor
\begin{equation}\label{eq:glm}
\eta_{sit}=o_{sit}+a_i+w_{it}(s).
\end{equation}
Here $o_{sit}$ is a known offset, for example log population, and $a_i$ is a variable-specific intercept with a vague Gaussian prior. The density of $Y_{sit}$ is $\exp[\{y\,\eta_{sit}-b_{sit}(\eta_{sit})\}/\varsigma+c(y,\varsigma)]$, with a known or estimated dispersion $\varsigma$ and a convex cumulant function $b_{sit}$. Examples are the Gaussian family with $b(\eta)=\eta^2/2$ and $\varsigma=\sigma^2$, which reproduces the BMW observation model when $T=1$, the Poisson family with $b(\eta)=e^{\eta}$, and the binomial family with $N_{sit}$ trials, $b(\eta)=N_{sit}\log(1+e^{\eta})$. The family enters only through the score $\{y-b'(\eta)\}/\varsigma$ and the weight $b''(\eta)/\varsigma$ of each response. Because $b$ is convex, the log-likelihood is concave in the latent fields, and the posterior mode of \S\ref{sec:laplace} is unique. The graph step, the switching rates and the parameter updates act on the latent innovations, so they are the same for every family. The likelihood omits missing responses.

\section{Structure of the process}\label{sec:structure}

Stack $w_{1:T}$ in time--variable--location order. In this section $S_T$ is the $T\times T$ lower shift matrix, and $G=I_{Tpn}-S_T\otimes R$, so that $\eta_{1:T}=Gw_{1:T}$. For each eigenmode $h$ and time $t$, the vector of mode coefficients is $\omega_{h,t}=(u_h^\top w_{1t},\dots,u_h^\top w_{pt})^\top\in\mathbb R^p$. We write $L_{h,t}$, $\tilde L_{h,t}$ and $Q_{h,t}$ for the matrices \eqref{eq:Lh} with $B$ replaced by $B_t$, and $\Phi_h=\diag\{r_1(\lambda_h),\dots,r_p(\lambda_h)\}$. The proofs of this section are in Supplement~\Sref{sec:proofs}.

\begin{theorem}[Eigenmode factorization]\label{thm:modes}
Under Assumption~\ref{ass:R}, the processes $\{\omega_{h,\cdot}\}_{h=1}^n$ are mutually independent, and each is a $p$-variate linear state-space model
\[
\omega_{h,t}=\Phi_h\,\omega_{h,t-1}+\xi_{h,t},\qquad \xi_{h,t}\sim N(0,Q_{h,t}^{-1})\ \text{independently over }t,\qquad \omega_{h,0}=0.
\]
\end{theorem}

\begin{theorem}[Log-determinant]\label{thm:logdet}
For every transition $R$, every sequence of graphs $\GV_1,\dots,\GV_T$ and every sequence of coefficients $B_1,\dots,B_T$, the precision matrix of $w_{1:T}$ satisfies
\[
\log|Q_{w(1:T)}|=\sum_{t=1}^T\log|Q_t|=T\Big\{\sum_{j=1}^p\alpha_j\sum_{h=1}^n\log(\kappa_j^2+\lambda_h)-2n\sum_{j=1}^p\log\tau_j\Big\}.
\]
\end{theorem}

The normalizing constant of the latent law is therefore the same closed-form function of $(\kappa,\alpha,\tau)$ whether the graph is empty, static or signed and switching, and whatever the dynamics. Proposition~\ref{prop:fb} relies on this property. Without it, the posterior over time-varying graphs would require one stochastic log-determinant for every configuration of the links. The result uses the initial condition $w_0=0$. Under a stationary initial law with link-dependent covariance, the graph-free property is lost (Supplement~\Sref{sec:proofs}).

\begin{theorem}[Markov structure]\label{thm:markov}
Suppose that $R=\blkdiag_i\{r_i(\Le)\}$ with $0<|r_i(\lambda)|<1$ for all $\lambda$. For $(t,i)\neq(s,j)$, and for Lebesgue-almost every value of the coupling coefficients, the $n\times n$ block of $Q_{w(1:T)}$ indexed by $\{(t,i),(s,j)\}$ is nonzero if and only if one of the following holds:
\begin{enumerate}
\item[(a)] $s=t$ and $i,j$ are adjacent in $\Mor(\GV_t)\cup\Mor(\GV_{t+1})$, with $\Mor(\GV_{T+1})$ taken to be empty, or
\item[(b)] $|s-t|=1$ and either $i=j$ or $i,j$ are adjacent in $\Mor(\GV_{\max(s,t)})$.
\end{enumerate}
Here $\Mor(\cdot)$ is the moral graph \citep{lauritzen1996}, which adds a link between any two parents of a common child. Condition~(b), and condition~(a) for a static graph, hold for every value of the coefficients.
\end{theorem}

Theorem~\ref{thm:markov} shows that the time-varying graph is a property of the process itself. In a Gaussian process, the zero blocks of the precision matrix are the conditional independences. The theorem identifies all nonzero blocks of $Q_{w(1:T)}$, and they follow from the graphs $\GV_1,\dots,\GV_T$ alone. If $i$ and $j$ are non-adjacent in the moral graphs at times $t$ and $t+1$, then $w_{it}$ and $w_{jt}$ are conditionally independent given all other fields at all times. An absent link is thus an exact conditional independence of the fields, localized in time, and not only a zero coefficient in a fitted model. Supplement~\Sref{sec:markovex} works through an example. For a static graph, the conditional-independence graph of $\{w_{it}\}$ is the strong product of the moral graph with the path on $T$ vertices.

\begin{corollary}[Exact Gaussian likelihood]\label{cor:kalman}
Suppose $Y_{sit}=w_{it}(s)+\varepsilon_{sit}$ with $\varepsilon_{sit}\sim N(0,\sigma^2)$, observed at all locations. Then the log-likelihood is the sum over $h$ of the Kalman-filter log-likelihoods of the rotated data $(u_h^\top Y_{\cdot 1t},\dots,u_h^\top Y_{\cdot pt})^\top=\omega_{h,t}+N(0,\sigma^2I_p)$, with the dynamics of Theorem~\ref{thm:modes}. After one eigendecomposition of $\Le$, it costs $O(nTp^3)$ operations.
\end{corollary}

\section{Inference}\label{sec:inference}

\subsection{Laplace approximation of the latent field}\label{sec:laplace}
All computations below use the spectral coordinates $\theta=\{\omega_{h,t}\}$ of the fields. By Theorem~\ref{thm:modes} the prior precision of $\theta$ is $Q_\theta=\cA^\top\cA$, where $\cA$ applies $\omega_{h,t}\mapsto\tilde L_{h,t}(\omega_{h,t}-\Phi_h\omega_{h,t-1})$ separately for each mode. For fixed parameters and graphs, the posterior mode $(\hat\theta,\hat a)$ maximizes the strictly concave function
\[
f(\theta,a)=\ell(\theta,a)-\tfrac12\theta^\top Q_\theta\theta-\tfrac12\gamma_a\|a\|^2,
\]
where $\ell$ is the log-likelihood and $\gamma_a$ the prior precision of the intercepts. Newton's method uses the negative Hessian
\begin{equation}\label{eq:P}
P=\begin{pmatrix}Q_\theta+H&H_{\theta a}\\H_{\theta a}^\top&H_{aa}+\gamma_aI\end{pmatrix},\qquad H=\blkdiag_{(i,t)}\big(U^\top W_{it}U\big),
\end{equation}
with $W_{it}=\diag\{b''_{sit}(\hat\eta_{sit})/\varsigma\}$ at the observed sites and zero elsewhere. The data term is block diagonal over variable--time pairs, because each response involves exactly one pair. The resulting Gaussian approximation is $q(\theta,a)=N\{(\hat\theta,\hat a),P^{-1}\}$. By Theorem~\ref{thm:logdet}, the Laplace approximation of the marginal likelihood contains the closed-form $\tfrac12\log|Q_\theta|$ and the term $-\tfrac12\log|P|$. The latter depends on the graph, and the algorithm below avoids its evaluation.

\begin{lemma}[Exact sampling from $q$]\label{lem:perturb}
Suppose that $z_1,z_2,z_3$ are independent standard normal vectors of the appropriate dimensions, that $X$ is the design that maps $\theta$ to the linear predictors of the observed responses, and that $W$ collects their weights. With
\[
u=\big(\cA^\top z_1+X^\top W^{1/2}z_2,\ \mathbf 1^\top W^{1/2}z_2+\gamma_a^{1/2}z_3\big),
\]
$\mathrm{cov}(u)=P$, and $(\hat\theta,\hat a)+P^{-1}u$ is an exact draw from $q$.
\end{lemma}

\subsection{The signed graph step}\label{sec:graphstep}\label{sec:mcem}
We write $\psi=(\kappa,\alpha,\tau,\phi)$ and $\mathcal B=(B_1,\dots,B_T)$. Given draws $\omega^{(1)},\dots,\omega^{(S)}$ from $q$, form the innovations $\xi^{(s)}_{h,t}=\omega^{(s)}_{h,t}-\Phi_h\omega^{(s)}_{h,t-1}$. A Monte Carlo expectation--maximization (EM) algorithm maximizes the Monte Carlo estimate of
\begin{equation}\label{eq:Qfun}
\mathcal Q(\psi,\mathcal B)=\E_q\log p(\omega\mid\psi,\mathcal B)=\tfrac12\sum_t\log|Q_t|-\tfrac12\sum_{h,t}\E_q\|\tilde L_{h,t}\,\xi_{h,t}\|^2 .
\end{equation}
Row $i$ of $\tilde L_{h,t}\xi_{h,t}$ is $\tau_i^{-1}\{d_i(\lambda_h)\xi_{h,t,i}+\sum_jb_{ij}(t)\xi_{h,t,j}\}$. By Theorem~\ref{thm:logdet}, \eqref{eq:Qfun} therefore splits over variables and times:
\begin{equation}\label{eq:Qsplit}
\mathcal Q=\sum_{i=1}^p\sum_{t=1}^T\Big[\sum_h\log d_i(\lambda_h)-n\log\tau_i-\frac{1}{2\tau_i^2}\E_q\sum_h\Big\{d_i(\lambda_h)\xi_{h,t,i}+\sum_jb_{ij}(t)\xi_{h,t,j}\Big\}^2\Big].
\end{equation}
For child $i$, candidate parents $\mathcal C_i=\{j:(i,j)\in\mathcal C\}$ and a parent set $\mathcal S\subseteq\mathcal C_i$, define the time-$t$ statistics
\[
\mathsf G_{i,t,\mathcal S}=\tau_i^{-2}\E_q\sum_h\xi_{h,t,\mathcal S}\xi_{h,t,\mathcal S}^\top,\qquad
\mathsf g_{i,t,\mathcal S}=-\tau_i^{-2}\E_q\sum_hd_i(\lambda_h)\,\xi_{h,t,i}\,\xi_{h,t,\mathcal S},
\]
estimated by averages over the draws. The $(i,t)$ term of \eqref{eq:Qsplit}, restricted to parent set $\mathcal S$, is $c_{i,t}+\mathsf g_{i,t,\mathcal S}^\top b-\tfrac12b^\top\mathsf G_{i,t,\mathcal S}b$ with $b=b_{i\mathcal S}(t)$, and $c_{i,t}$ does not depend on $\mathcal S$. A sign configuration $c\in\{-1,0,+1\}^{\mathcal C_i}$ of the links of child $i$ fixes the parent set $\mathcal S=\{j:c_j\neq0\}$ and the mean coefficients $\mu_c$, with entries $-\beta^+_{ij}$ for $c_j=+1$ and $\beta^-_{ij}$ for $c_j=-1$. Integration of the deviations $\delta$ of \eqref{eq:edgecoef} against the $(i,t)$ term gives the emission, the contribution of configuration $c$ at time $t$ to the likelihood of the graph,
\[
e_{i,t}(c;\beta)=-\tfrac12\log\big|I+s_\delta^2\mathsf G_{i,t,\mathcal S}\big|+\tfrac12 v^\top\big(\mathsf G_{i,t,\mathcal S}+s_\delta^{-2}I\big)^{-1}v-\frac{\|\mu_c\|^2}{2s_\delta^2},\qquad v=\mathsf g_{i,t,\mathcal S}+s_\delta^{-2}\mu_c.
\]
We treat $\exp(\mathcal Q)$ as the likelihood of $\mathcal B$ given the E-step, and the strengths $\beta$ as parameters with half-normal penalties.

\begin{proposition}[Exact signed graph posterior]\label{prop:fb}
For a fixed variable ordering and given $\beta$, the posterior over the link states $\{\gamma_{ij}(t)\}$ factorizes over children $i$. For each child it is the posterior of a hidden Markov model with the following components:
\begin{itemize}
\item state space the joint augmented states of the chains \eqref{eq:edgechain} of the candidate parents, $8^{|\mathcal C_i|}$ states,
\item initial law and transition matrix the Kronecker products of those of the link chains,
\item emissions $\exp\{e_{i,t}(c;\beta)\}$, which depend on the joint state only through the current sign configuration $c$.
\end{itemize}
The forward--backward algorithm computes its marginals, and so the probabilities $p^\pm_{ij}(t)=\Pr\{\gamma_{ij}(t)=\pm1\}$, exactly, in $O(T|\mathcal C_i|\,8^{|\mathcal C_i|+1})$ operations plus $O(T\,3^{|\mathcal C_i|}|\mathcal C_i|^3)$ for the emissions.
\end{proposition}

Because $\beta$ is plugged in rather than integrated out, the emissions lack the Occam factor of a marginal likelihood. This factor is the penalty that integration over the prior of a parameter, here a strength, places on a model with that parameter. We restore it approximately by the log Bayes factor $o_{ij}=-\tfrac12\log(1+s^2\sum_t\mathsf G_{i,t,jj})$, applied at the first use of each sign: the transition into a sign not used before is multiplied by $e^{o_{ij}}$. A return to a sign already used receives no penalty, because its strength is shared over periods and its penalty has already entered. The penalty multiplies paths, so the posterior remains an exact hidden Markov model. With $\epsilon_{\rm off}=0$ it reproduces the static inclusion odds to first order.

Given the posterior, the strengths maximize the expected complete-data objective in the limit $s_\delta\to0$, subject to $\beta^\pm_{ij}\ge0$. For child $i$ this is a bounded least-squares problem in $2|\mathcal C_i|$ strengths (Supplement~\Sref{sec:chain}). Ten alternations of forward--backward recursions and strength updates give the graph posterior. Ten alternations suffice. In two simulated Poisson data sets of \S\ref{sec:sim}, one with and one without sign changes, the strengths after ten alternations differ from those after forty by less than 2\% of the largest strength, and the inclusion probabilities by less than 0.002 (Supplement~\Sref{sec:chain}). A sign whose strength reaches its bound at zero is removed from the state space of its link.

The inclusion probability of link $(i,j)$ at time $t$ is $p_{ij}(t)=p^+_{ij}(t)+p^-_{ij}(t)$. A graph is selected by the Bayesian false discovery rate rule \citep{newton2004}. For a target level $q$, this rule selects the largest set of (link, time) pairs with the highest inclusion probabilities whose mean posterior probability of absence is at most $q$. The expected share of false links among the selected pairs, under the posterior, is then at most $q$. Each selected pair takes the sign of larger posterior probability. To keep the forward--backward computation small, each variable has at most five candidate parents. They are the five variables earlier in the ordering with the largest partial correlations with it, computed from the innovations of a fit without a graph, each scaled to unit variance. With $\epsilon_{\rm off}=0$ the posterior over parent sets is available in closed form, with the coefficients integrated out exactly (Supplement~\Sref{sec:proofs}).

\subsection{Switching rates with an interval}\label{sec:rates}
The rates are estimated from a fit without a graph, so that no fitted graph can influence the evidence for change. The step has four parts (details in Supplement~\Sref{sec:ratestep}).
\begin{enumerate}
\item From $S'=20$ draws of the fit without a graph, regression calibration \citep{carroll2006} removes the prior variance of each innovation and gives a pseudo-observation of it with a known error variance.
\item For each child $i$ and time $t$, a weighted regression over the eigenmodes of the pseudo-observation of $i$ on the innovations of its candidate parents gives coefficients $\hat b_{ij}(t)$ with working standard errors $s_{ij}(t)$. The candidates of this step are the parents with $|z|>3$ in a regression pooled over periods.
\item The working standard errors understate the variability of $\hat b_{ij}(t)$, by a factor specific to the child when the link is absent and by a further factor when it is present. The model
\begin{equation}\label{eq:calemission}
\hat b_{ij}(t)\mid\gamma_{ij}(t)=0\ \sim\ N\{0,\,c_{0,i}^2s_{ij}(t)^2\},\qquad \hat b_{ij}(t)\mid\gamma_{ij}(t)\neq0\ \sim\ N\{\beta_{ij},\,c_1^2s_{ij}(t)^2\},
\end{equation}
with $\beta_{ij}\sim N(0,s^2)$, a two-state version of the chain \eqref{eq:edgechain}, and a share $\varrho$ of candidates that are links at all, is fitted by EM. Its E-step is a forward--backward recursion for each candidate, and its M-step is in closed form.
\item The 95\% intervals for $\epsilon_{\rm off}$ and $\epsilon_{\rm on}$ are Wald intervals on the logit scale, from the observed information at the maximum.
\end{enumerate}
The step uses on/off states with one strength per link, so a link that changes sign counts as present in both phases. In the simulations with sign changes, the estimate falls slightly below the realized rate (\S\ref{sec:simresults}).

\subsection{Parameters and computation}\label{sec:computation}
Under scalar persistence, the terms of \eqref{eq:Qfun} with $t\ge2$ are quadratic in $\phi$, and the persistence update solves a $p\times p$ linear system. For each $i$, \eqref{eq:Qsplit} is maximized over $\tau_i$ in closed form, and the profile is then maximized over $\log\kappa_i$ by a bounded one-dimensional search. The log-determinant term makes different values of $\alpha_i$ comparable, so the same profile selects $\alpha_i$ on a grid. For the Gaussian family, $\sigma^2$ is updated from paired draws of fields and intercepts. An alternation of the Laplace approximation, the draws, the graph step and these updates gives a Laplace--Monte Carlo EM algorithm (Supplement Algorithm~\ref{alg:mcem}).

Every Newton step and every draw requires a solve with $P$. Preconditioned conjugate gradients carry out these solves without forming $P$, and with a block preconditioner and deflation they take a few tens of iterations (Supplement~\Sref{sec:matfree}).

\subsection{The DBMW estimator}\label{sec:procedure}\label{sec:estimators}
\begin{figure}[!htbp]
\centering
\definecolor{posc}{HTML}{2A78D6}\definecolor{negc}{HTML}{EB6834}
\begin{tikzpicture}[x=0.95cm,y=0.95cm,
  tile/.style={inner sep=0pt,draw=black!40,line width=0.3pt,fill=white},
  vnode/.style={circle,draw=black!70,line width=0.5pt,minimum size=4.2mm,inner sep=0pt,font=\scriptsize},
  hnode/.style={circle,draw=black!60,line width=0.4pt,minimum size=3.2mm,inner sep=0pt,fill=white},
  gen/.style={-{Latex[length=2.2mm]},line width=0.7pt,draw=black!60},
  est/.style={-{Latex[length=2.2mm]},line width=1.0pt,draw=orange!80!black},
  inner/.style={-{Latex[length=1.8mm]},line width=0.6pt,draw=black!55},
  stage/.style={rounded corners=3pt,line width=0.8pt},
  lab/.style={font=\footnotesize,align=center},
  on/.style={-{Latex[length=1.6mm]},line width=0.8pt,draw=negc},
  off/.style={line width=0.5pt,draw=black!30,dash pattern=on 1.2pt off 1.2pt}]
\def\st#1#2#3{\foreach \k/\f in {2/#3_3,1/#3_2,0/#3_1}{ \node[tile] at (#1+0.14*\k,#2+0.14*\k) {\includegraphics[width=1.1cm]{tiles/\f.png}}; } }
\node[anchor=west,font=\small\itshape] at (-0.2,8.15) {Model};
\foreach \y/\p in {7.45/{-1,-1,-1,0,0,1},7.15/{0,0,1,1,1,1},6.85/{0,0,0,0,0,0}}{
  \foreach \v [count=\k] in \p { \pgfmathsetmacro{\xx}{0.3+0.3*(\k-1)}
    \ifnum\v=1 \fill[posc!70] (\xx,\y-0.1) rectangle (\xx+0.28,\y+0.1); \else\ifnum\v=-1 \fill[negc!80] (\xx,\y-0.1) rectangle (\xx+0.28,\y+0.1); \else \draw[black!30,line width=0.4pt] (\xx,\y-0.1) rectangle (\xx+0.28,\y+0.1); \fi\fi } }
\node (gam) at (1.2,7.15) [minimum width=2cm,minimum height=1.1cm] {};
\node[lab] at (1.2,5.8) {Link states\\ $\gamma_{ij}(t)$};
\node[vnode] (n1) at (3.6,7.65) {1}; \node[vnode] (n2) at (4.6,7.65) {2}; \node[vnode] (n3) at (4.1,6.75) {3};
\draw[on] (n1)--(n3); \draw[off] (n2)--(n3);
\node (dag) at (4.1,7.2) [minimum width=1.6cm,minimum height=1.3cm] {};
\node[lab] at (4.1,5.8) {Graph\\ $\mathcal G_t,\ B_t$};
\st{6.75}{7.0}{w2} \node (xi) at (6.9,7.15) [minimum width=1.5cm,minimum height=1.5cm] {};
\node[lab] at (6.9,5.8) {Innovations\\ $\xi_t$};
\st{9.75}{7.0}{w1} \node (ww) at (9.9,7.15) [minimum width=1.5cm,minimum height=1.5cm] {};
\node[lab] at (9.9,5.8) {Fields\\ $w_t=Rw_{t-1}+\xi_t$};
\foreach \k in {2,1,0}{ \node[tile] at (12.95+0.14*\k,7.0+0.14*\k) {\includegraphics[width=1.1cm]{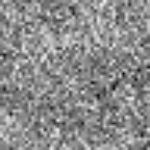}}; }
\node (yy) at (13.1,7.15) [minimum width=1.5cm,minimum height=1.5cm] {};
\node[lab] at (13.1,5.8) {Responses\\ $Y_{sit}$};
\node[inner sep=0pt] (sites) at (6.9,9.05) {\includegraphics[width=0.85cm]{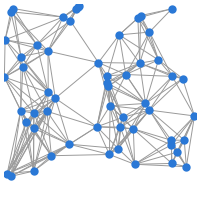}};
\node[lab,anchor=east] at (6.35,9.05) {Sites and\\ neighbour graph, $L_\varepsilon$};
\draw[gen] (sites.south)--(6.9,7.95);
\draw[gen] (gam.east)--(dag.west); \draw[gen] (dag.east)--(xi.west); \draw[gen] (xi.east)--(ww.west); \draw[gen] (ww.east)--(yy.west);
\draw[gen] (ww.north) to[out=110,in=70,looseness=4] node[above,font=\footnotesize]{$R$} (ww.north east);
\node[anchor=west,font=\small\itshape] at (-0.2,5.17) {Estimation};
\draw[stage,draw=blue!60!black,fill=blue!4] (0.0,2.55) rectangle (16.3,4.85);
\node[anchor=north west,font=\small] at (0.1,4.8) {\textbf{Step 1}\enspace Fit without a graph};
\foreach \k/\x in {1/1.3,2/2.05,3/2.8}{ \node[tile] at (\x,3.55) {\includegraphics[width=0.62cm]{tiles/eig\k.png}}; }
\node (modes) at (2.05,3.55) [minimum width=2.3cm,minimum height=0.7cm] {};
\node[lab] at (2.05,2.85) {Spatial eigenmodes of $L_\varepsilon$};
\st{6.6}{3.65}{w3} \node (draws) at (6.75,3.8) [minimum width=1.5cm,minimum height=1.4cm] {};
\node[lab] at (6.75,2.85) {Posterior draws of $\xi_t$};
\draw[black!60,line width=0.6pt] (10.6,3.55)--(13.4,3.55); \draw[line width=1.6pt,blue!60!black] (11.2,3.55)--(12.3,3.55); \fill[blue!60!black] (11.65,3.55) circle (2.2pt);
\node (rates) at (12.0,3.55) [minimum width=2.8cm,minimum height=0.5cm] {};
\node[lab] at (12.0,2.95) {Switching rates $\hat\epsilon_{\rm on},\hat\epsilon_{\rm off}$\\ with 95\% interval};
\draw[inner] (modes.east)--(5.9,3.55); \draw[inner] (7.65,3.55)--(rates.west);
\draw[est] (13.1,5.3) -- (13.1,4.85);
\draw[stage,draw=green!45!black,fill=green!4] (0.0,0.0) rectangle (7.9,2.05);
\node[anchor=north west,font=\small] at (0.1,2.0) {\textbf{Step 2}\enspace Exact graph posterior};
\foreach \k in {1,...,5}{ \node[hnode] (h\k) at (0.15+0.55*\k,1.05) {}; }
\foreach \a/\b in {1/2,2/3,3/4,4/5}{ \draw[-{Latex[length=1.3mm]},black!60,line width=0.5pt] (h\a)--(h\b); }
\node[lab] at (1.8,0.4) {Forward--backward\\ per variable};
\node[inner sep=0pt] (pip) at (5.6,1.05) {\includegraphics[width=2.6cm,height=0.38cm]{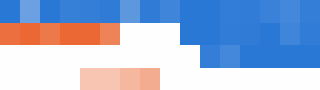}};
\node[lab] at (5.6,0.4) {Inclusion probabilities\\ and signs};
\draw[inner] (h5.east)--(pip.west);
\node[anchor=north east,font=\footnotesize\itshape] at (7.9,-0.08) {DBMW-X ends here};
\draw[stage,draw=orange!80!black,fill=orange!5] (8.4,0.0) rectangle (16.3,2.05);
\node[anchor=north west,font=\small] at (8.5,2.0) {\textbf{Step 3}\enspace Joint refit};
\node[lab] (fld) at (9.7,1.0) {Fields $w$};
\node[lab] (grs) at (12.1,1.0) {Graph and\\ strengths};
\draw[-{Latex[length=1.8mm]},line width=0.7pt,orange!80!black] (fld.north east) to[bend left=30] (grs.north west);
\draw[-{Latex[length=1.8mm]},line width=0.7pt,orange!80!black] (grs.south west) to[bend left=30] (fld.south east);
\node[lab] (outp) at (14.85,1.0) {Strengths, fields,\\ predictions};
\draw[inner] (grs.east)--(outp.west);
\node[anchor=north east,font=\footnotesize\itshape] at (16.3,-0.08) {DBMW ends here};
\draw[est] (6.75,2.55) -- (6.75,2.05);
\draw[est] (7.9,0.7) -- (8.4,0.7);
\end{tikzpicture}
\caption{Architecture of the model and of the estimator. Top: the model, from the link states through the graph and the innovations to the fields and the responses. The innovations are spatial fields built on the neighbour graph of the sites, through its Laplacian $\Le$. Bottom: the three steps of the estimator. Step~1 fits the fields without a graph, in the eigenmodes of $\Le$, and estimates the switching rates from posterior draws of the innovations. Step~2 computes the exact posterior of the signed graph. Step~3 refits fields, graph and strengths jointly.}\label{fig:pipeline}
\end{figure}

Figure~\ref{fig:pipeline} and Algorithm~\ref{alg:full} combine these blocks into the DBMW estimator, in three steps.

In Step~1, the Laplace--Monte Carlo EM algorithm with no graph fits one field per variable, with smoothness $\alpha_i=2$. Counts carry little information about smoothness, so the first fit fixes it, and later rounds choose it on a grid. At the final parameters, $S'=20$ draws give the innovations of every variable at every time. The rate step of \S\ref{sec:rates} uses these draws. A fit without a graph favours no link, so no fitted graph influences the evidence for change. Four rounds of link-aware updates then refine $(\kappa_i,\alpha_i,\tau_i,\phi_i)$, with pooled coefficients for the rate candidates and $\alpha_i$ on a grid. These rounds correct the parameters of the fit without a graph. Without a graph, the model attributes the contribution of a variable's parents to the variable's own field, which inflates its amplitude $\tau_i$ and distorts its range $\kappa_i$.

In Step~2, the signed graph step of \S\ref{sec:graphstep} runs once, on the Step-1 draws and at the Step-1 rates. Its statistics use the regression-calibrated innovations of the rate step. The scale $\tau_i$ of each child is multiplied by $\max(1,\hat c_{0,i})$, the calibration factor of \eqref{eq:calemission}. This factor widens the noise scale when the per-period coefficients vary more than their working standard errors, and it never narrows the scale. A rate rule then checks the evidence for switching. If the lower end of the interval for $\epsilon_{\rm off}$ lies below $10^{-3}$, the static posterior of DBMW-0 replaces the signed dynamic one. DBMW-X ends here, with the conditional posterior mode of the fields given the Step-2 graph.

In Step~3, the Laplace--Monte Carlo EM algorithm fits fields, graph and parameters jointly, with the signed graph step in every iteration. It starts from the Step-2 graph and strengths and from the link-aware parameters, and it keeps the rates and the smoothness at their Step-1 values. The refit reduces the attenuation of the strengths, because each field now borrows information from its parents, and it gives better predictions of withheld data. Steps 1--3 together form DBMW.

Each step answers its own question. The graph comes from the Step-2 inclusion probabilities. Given the Step-1 parameters, they form an exact empirical Bayes posterior, close to calibrated in the simulations. A joint fit of graph and fields from the start makes them overconfident, because the prior coupling of an included link feeds back into its own evidence (Supplement~\Sref{sec:choice}). Strengths, fields and predictions come from Step~3. The switching rates and their intervals come from Step~1. Predictions use the posterior mean of the expected response, from 20 draws of the Laplace approximation.

\begin{algorithm}[!htbp]
\caption{The DBMW estimator}\label{alg:full}
\begin{algorithmic}[1]
\Require responses $Y_{sit}$ with missingness indicators, site coordinates, a variable ordering, the family and offsets
\State Build the 10-nearest-neighbour graph of the sites, the density-normalized Laplacian $\Le$ with unit mean diagonal and its eigenpairs $(\lambda_h,u_h)$.
\Statex \emph{Step 1: fields without a graph}
\State Fit $(\kappa_i,\tau_i,\phi_i)$, the intercepts and, for the Gaussian family, $\sigma^2$ by Laplace--Monte Carlo EM with no graph and $\alpha_i=2$.
\State Compute the posterior mode and $S'=20$ draws of the fields from $q$ (Lemma~\ref{lem:perturb}), with their innovations.
\State Estimate $\epsilon_{\rm on}$ and $\epsilon_{\rm off}$ with 95\% intervals, and the calibration factors $c_{0,i}$, by the rate step of \S\ref{sec:rates}.
\State Refine $(\kappa_i,\alpha_i,\tau_i)$, with $\alpha_i$ on a grid, and $\phi$ in four rounds with pooled coefficients for the rate candidates.
\State Take all pairs as candidates for small $p$, and otherwise five earlier-ordered partners per variable from the partial-correlation screen.
\Statex \emph{Step 2: exact graph posterior}
\State Compute the signed graph posterior from the calibrated statistics of the draws of line 3, with scales $\tau_i\max(1,\hat c_{0,i})$, by ten alternations of forward--backward recursions (Proposition~\ref{prop:fb}) and strength updates, at $(\hat\epsilon_{\rm on},\hat\epsilon_{\rm off})$ with $\pi=0.2$.
\State Use the static graph of DBMW-0 instead if the lower end of the $\epsilon_{\rm off}$ interval lies below $10^{-3}$.
\State Compute the conditional mode of the fields given the Step-2 graph, which ends DBMW-X.
\Statex \emph{Step 3: joint refit}
\State Refit fields, graph and parameters by Laplace--Monte Carlo EM with the signed graph step, from the Step-2 graph and the parameters of line 5, at the rates of line 4.
\State Report the Step-2 inclusion probabilities and signs with Bayesian false discovery selection, the switching rates with intervals, and the Step-3 strengths, fields and predictions.
\end{algorithmic}
\end{algorithm}

\section{Simulation study}\label{sec:sim}
The simulation study uses data sets with known fields, graph and link paths. It addresses three questions. The first concerns the recovery of the links, their times and their signs, against established graph estimators. The second concerns the calibration of the inclusion probabilities and the switching rates. The third concerns prediction, against a spatio-temporal model that fits each variable separately.

\subsection{Design}\label{sec:simdesign}
The locations are $n=200$ points drawn uniformly on the unit square. The true $\Le$ is the density-normalized Laplacian of the 15-nearest-neighbour graph, and every fit uses the 10-nearest-neighbour graph of the coordinates instead. There are $p=20$ variables, with independent parameters
\[
\kappa_i\sim\mathrm{Un}(0.25,0.45),\qquad \alpha_i\sim\mathrm{Un}(1.6,2.8),\qquad \tau_i\sim\mathrm{Un}(0.5,2.0),\qquad \phi_i\sim\mathrm{Un}(0.5,0.9),
\]
and scalar persistence. Chain links $i\leftarrow i-1$ are present with probability $0.6$, and longer-range links $i\leftarrow i-l$, $l=2,3,4$, with probability $0.15$, about 20 links in total. A link has coefficient $\pm\rho_{ij}\kappa_i^{\alpha_i}$ with $\rho_{ij}\sim\mathrm{Un}(0.7,0.95)$, a scaling that keeps $|b_{ij}|<d_i(\lambda_h)$ in every mode. Every true link follows a two-state chain. It is present at $t=1$ with probability $1/2$ and changes state with probability $\epsilon$ at each later time. The fields come exactly from the model, one eigenmode at a time (Theorem~\ref{thm:modes}). A random 10\% of the (variable, time) maps is withheld for prediction.

The main text reports four scenarios.
\begin{itemize}
\item S1G, S1P and S1N have $T=26$ and $\epsilon=0.1$, about 2.5 switches per link, with a fixed sign for each link. S1G has Gaussian responses with noise variance $\mathrm{var}(w)/10$. S1P has Poisson counts, $Y_{sit}\sim\text{Poisson}[3\exp\{w_{it}(s)\}]$, after a common factor scales the fields to unit standard deviation. S1N has binomial counts with $N=10$ trials, logit link and baseline probability $0.3$.
\item S1PF has Poisson counts with sign changes, $T=52$ and $\epsilon=0.05$. Each time a link enters, it draws a new random sign, so a link alternates between phases of about 20 periods of either sign or of absence, as in the air-quality data.
\end{itemize}
Supplement~\Sref{sec:other} reports eight further scenarios: Gaussian responses with sign changes, binary responses, negative binomial counts, link-specific rates, gradual switches, a static graph with Gaussian and with Poisson responses, and a shared spatio-temporal component without a graph. Ten data sets per scenario use new locations, parameters, graph, link paths and responses.

\subsection{Methods compared}\label{sec:simmethods}
We fit DBMW and DBMW-X by Algorithm~\ref{alg:full}, with 15 outer iterations, $S=5$ draws per iteration, $S'=20$ draws for the rates and the graph, $\pi=0.2$, $s=0.5$ and $s_\delta=0.05$. The candidate screen keeps 85 of the 190 pairs.

Four established graph estimators score all 190 pairs in every period.
\begin{itemize}
\item GL is the graphical lasso \citep{friedman2008} with threefold cross-validation, applied to the innovations of the fit without a graph of Step~1, each scaled to unit variance, with one static graph.
\item GL-W applies the same estimator in a moving window of five periods.
\item TVGL is the time-varying graphical lasso \citep{hallac2017} on the same innovations, with a Laplacian penalty on changes and both penalties chosen by held-out likelihood.
\item PLN is the Poisson log-normal network \citep{chiquet2019} on the counts, with the penalty chosen by StARS \citep{liu2010}, for Poisson responses.
\end{itemize}
Each estimator scores a pair by the absolute partial correlation and selects it when the precision entry is nonzero. On the standardized responses instead of the innovations, GL and TVGL recover the graph less well (Supplement~\Sref{sec:other}).
For prediction, the competitor SPDE-AR1 fits one Mat\'ern SPDE field per variable with AR(1) evolution \citep{cameletti2013}, by R-INLA \citep{rue2009, lindgren2011}, with penalized-complexity priors and the default mesh of Supplement~\Sref{sec:competitors}. All methods score the withheld entries at the posterior mean of the expected response. The loss is the mean squared error for Gaussian responses and the mean deviance otherwise.

\subsection{Results}\label{sec:simresults}
\begin{figure}[!htbp]
\centering
\includegraphics[width=\linewidth]{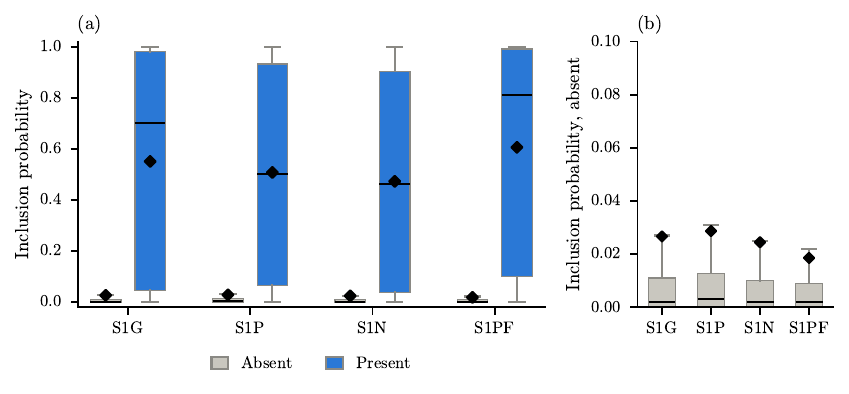}
\caption{Inclusion probabilities of DBMW for absent and present (link, period) pairs, pooled over ten data sets per scenario. (a) All pairs. (b) Absent pairs only, on a finer scale. Boxes: quartiles. Whiskers: 1.5 interquartile ranges. Diamonds: means.}\label{fig:pip}
\end{figure}

The truth $\gamma_{ij}(t)$ is known for every pair and period, with about ten present links among the 190 pairs in each period. Table~\ref{tab:graph} gives the average precision (AP) of the scores and the false links per period of the selected graph. DBMW has the highest AP in all four scenarios, 0.63--0.71, with an area under the ROC curve of 0.89--0.92. It selects one false link per period or fewer. GL reaches an AP of 0.49--0.54 with a fixed sign, but it selects 24--26 false links per period. With sign changes (S1PF), its AP falls to 0.29, because one static graph averages the two signs of a link. The windowed version gains over GL only with sign changes. TVGL selects about 170 of the 190 pairs in every period. PLN uses all counts, including the withheld ones, and still reaches an AP of only 0.34 and 0.25. Figure~\ref{fig:pip} shows the inclusion probabilities of DBMW. Absent links have probabilities near zero, with a mean of 0.02--0.03. Present links have a median probability of 0.46--0.81, and a quarter of them stay below 0.1. The errors of DBMW are thus mostly missed links rather than false ones.

\begin{table}[t]
\centering
\caption{Graph recovery, means over ten data sets: average precision (AP) of the scores over all variable pairs and periods, and false links per period (FP) of the selected graph. DBMW: inclusion probabilities, selection at probability 0.5. GL: graphical lasso on the innovations of the fit without a graph. GL-W: the same in a moving window of five periods. TVGL: time-varying graphical lasso on the same innovations. PLN: Poisson log-normal network, a model for Poisson counts, so not applicable (--) to Gaussian and binomial responses. The best AP in each row is in bold.}\label{tab:graph}
\footnotesize
\begin{tabular}{lcccccccccc}
\toprule
& \multicolumn{2}{c}{DBMW} & \multicolumn{2}{c}{GL} & \multicolumn{2}{c}{GL-W} & \multicolumn{2}{c}{TVGL} & \multicolumn{2}{c}{PLN} \\
\cmidrule(lr){2-3}\cmidrule(lr){4-5}\cmidrule(lr){6-7}\cmidrule(lr){8-9}\cmidrule(lr){10-11}
Scenario & AP & FP & AP & FP & AP & FP & AP & FP & AP & FP \\
\midrule
S1G & \textbf{0.67} & 1.0 & 0.54 & 25.4 & 0.43 & 20.7 & 0.33 & 171.2 & -- & -- \\
S1P & \textbf{0.68} & 1.0 & 0.52 & 26.2 & 0.36 & 23.4 & 0.27 & 171.9 & 0.34 & 15.2 \\
S1N & \textbf{0.63} & 0.8 & 0.49 & 24.4 & 0.38 & 27.8 & 0.27 & 172.4 & -- & -- \\
S1PF & \textbf{0.71} & 0.6 & 0.29 & 4.5 & 0.37 & 17.3 & 0.27 & 172.7 & 0.25 & 8.7 \\
\bottomrule
\end{tabular}
\end{table}

Table~\ref{tab:unc} reports the uncertainty measures. The expected calibration error of the inclusion probabilities is 0.03--0.04. With the Bayesian false discovery rate rule of \S\ref{sec:graphstep} at $q=0.10$, the realized false discovery rate is 0.10--0.16. Selected pairs carry the correct sign in at least 99.9\% of cases, and no link with a single sign in the truth is selected with both signs. To measure the detection of links over time, we split the true path of each link into sign phases. A sign phase is a maximal run of consecutive periods in which the link is present with the same sign. For example, a link that is positive in periods 3--10, absent in periods 11--20 and negative in periods 21--30 has two sign phases, of lengths 8 and 10. A phase counts as found when the selected graph contains the link, with the true sign, in at least one period of the phase. Table~\ref{tab:unc} gives the share of phases found, by phase length. Detection rises with the length of the phase, from 21--40\% for phases of one to seven periods to 45--67\% for phases of 16 periods or more. With sign changes (S1PF), short phases are found less often and long phases more often than in S1P. A phase of a link whose truth has both signs is found less often than a phase of the same length of a link with one sign, for example in 32\% against 55\% of the phases of 8--15 periods (Supplement~\Sref{sec:other}). Both phases of a link that changes sign are found for only 9\% of such links, and for 15\% after the joint refit of Step~3. The power to detect a change of sign is thus limited by the evidence within each phase. A detected change of sign is reliable, because changes are almost never selected for links with one sign. The interval for $\epsilon_{\rm off}$ covers the realized rate in all data sets. With sign changes, the estimate falls slightly below the realized rate, 0.056 against 0.061 in S1PF, as expected from the on/off rate step.

\begin{table}[t]
\centering
\caption{Uncertainty, means over ten data sets. ECE: expected calibration error of the inclusion probabilities (ten bins). FDR: realized false discovery rate of the Bayesian rule at $q=0.10$. Correct: share of selected (link, period) pairs with the true sign. False flips: share of single-sign links selected with both signs. Phases found: share of true sign phases, by length in periods, in which the selected graph contains the link with the true sign in at least one period. $\hat\epsilon_{\rm off}$: estimated switching rate, with the realized rate in brackets. Coverage: share of 95\% intervals that contain the realized rate. Bold: the best value in each column, unless all values are equal. For FDR the best value is the closest to the nominal 0.10, and for $\hat\epsilon_{\rm off}$ the closest to the realized rate in relative terms.}\label{tab:unc}
\footnotesize
\begin{tabular}{lccccccccc}
\toprule
& \multicolumn{2}{c}{Inclusion} & \multicolumn{2}{c}{Signs} & \multicolumn{3}{c}{Phases found, by length} & \multicolumn{2}{c}{Switching rate} \\
\cmidrule(lr){2-3}\cmidrule(lr){4-5}\cmidrule(lr){6-8}\cmidrule(lr){9-10}
Scenario & ECE & FDR & Correct & False flips & 1--7 & 8--15 & $\geq 16$ & $\hat\epsilon_{\rm off}$ (true) & Coverage \\
\midrule
S1G & 0.039 & 0.15 & \textbf{1.000} & 0.000 & \textbf{0.40} & \textbf{0.57} & 0.55 & 0.093 (0.099) & 1.0 \\
S1P & 0.039 & 0.16 & \textbf{1.000} & 0.000 & 0.33 & 0.54 & 0.47 & 0.105 (0.099) & 1.0 \\
S1N & 0.042 & 0.13 & \textbf{1.000} & 0.000 & 0.29 & 0.49 & 0.45 & \textbf{0.101 (0.099)} & 1.0 \\
S1PF & \textbf{0.027} & \textbf{0.10} & 0.999 & 0.000 & 0.21 & 0.43 & \textbf{0.67} & 0.056 (0.061) & 1.0 \\
\bottomrule
\end{tabular}
\end{table}

On selected true links, the DBMW strengths have the correct sign and a correlation of 0.94--0.98 with the truth. They are attenuated, with a median relative error of $-11\%$ to $-23\%$. The DBMW-X strengths are more attenuated, by $-36\%$ to $-49\%$, which motivates the joint refit. The scale of $b_{ij}$ moves with the weakly identified $\kappa_i^{\alpha_i}$ \citep{zhang2004}, and the 95\% intervals for the strengths cover the truth in only 27--49\% of cases. We therefore report the sign and the point estimate of each strength.

Table~\ref{tab:pred} gives the loss on whole withheld maps. DBMW has the lowest loss in all four scenarios, 10--29\% below SPDE-AR1, and beats SPDE-AR1 in 39 of the 40 data sets. DBMW-X lies between the two. The gain of the joint refit over DBMW-X is largest for Poisson counts (S1P, 12\%). DBMW-X takes 6--22 minutes per data set on two cores, and DBMW 11--37 minutes.

\begin{table}[t]
\centering
\caption{Prediction of whole withheld maps and run time. Loss on the withheld entries (squared error for Gaussian data, deviance otherwise) at the posterior mean of the expected response, means over ten data sets. Wins: data sets in which DBMW has a lower loss than SPDE-AR1. Minutes per data set on two cores, means over three data sets.}\label{tab:pred}
\footnotesize
\begin{tabular}{lccccrr}
\toprule
& \multicolumn{3}{c}{Held-out loss} & & \multicolumn{2}{c}{Minutes} \\
\cmidrule(lr){2-4}\cmidrule(lr){6-7}
Scenario & SPDE-AR1 & DBMW-X & DBMW & Wins & DBMW-X & DBMW \\
\midrule
S1G & 7.99 & 6.36 & \textbf{5.98} & 10/10 & 6 & 11 \\
S1P & 4.18 & 3.37 & \textbf{2.96} & 10/10 & 11 & 21 \\
S1N & 1.897 & 1.742 & \textbf{1.706} & 10/10 & 12 & 22 \\
S1PF & 4.30 & 3.88 & \textbf{3.77} & 9/10 & 22 & 37 \\
\bottomrule
\end{tabular}
\end{table}

Supplement~\Sref{sec:other} reports the other scenarios. DBMW keeps the highest AP with Gaussian sign changes, 0.72 against 0.43 for the best baseline. Binary responses carry little information about the graph, and DBMW and GL reach an AP of only 0.43 and 0.44. Overdispersion raises the realized false discovery rate to 0.32 at a nominal 0.10. Under a static truth, GL matches DBMW in AP but selects 42--45 false links per period, and the rate rule chooses the static graph in 50\% and 20\% of the data sets. In one further study, one variable nearly vanishes in the middle third of the periods, as influenza did in 2020--22. With the drop in the offset, the inclusion probabilities of its true links stay at the same level inside and outside the window, so the vanishing variable produces no spurious switch-off. Without the drop in the offset, the probabilities fall at all times (Supplement~\Sref{sec:holdout}). A rate study over $n$, $T$, $p$ and $\epsilon$ gives intervals with 70--100\% coverage (Supplement~\Sref{sec:ratestudy}).

\section{Applications}\label{sec:applications}
Two applications illustrate the two main response types. The air-quality data are continuous, and the question concerns a known seasonal change of sign in the pollutant graph. The surveillance data are counts, and the question concerns the stability of the dependence between diseases through the COVID-19 pandemic. Both analyses use DBMW with all pairs as candidates and report the Step-2 graph with Bayesian false discovery selection at $q=0.10$. The directed graph depends on the variable order. Step~1 has no graph and does not depend on the order, so only the rate step and Step~2 are repeated for each of 120 orders. Each analysis reports undirected links, with the share of orders that select each link with each sign. Prediction uses two hold-out designs, fixed before any fit. \emph{Whole maps} withholds 10\% of the (variable, period) maps at all sites. \emph{Regions} splits the sites into ten spatial regions by $k$-means and withholds, for each variable, one region at all periods, with three such data sets. Supplement~\Sref{sec:apps} gives further details of the data and the fits.

\subsection{Air pollution across the United States}\label{sec:aqs}
The daily summary files of the Air Quality System of the U.S. Environmental Protection Agency \citep{epa_aqs} give the daily values of each monitor. We kept the sites in the contiguous United States with both ozone and \NOt\ in at least 80\% of the weeks of 2023, $n=288$. For each site and week, the response is the mean of the daily values over all monitors of the site, with at least four days of data. The variables are temperature, \NOt, CO, \PMt\ and ozone, with the pollutants on the log scale. Each variable is standardized over sites and weeks, and the family is Gaussian. \NOt\ and ozone are observed in 98\% of the site-weeks, and temperature, \PMt\ and CO in 79\%, 61\% and 34\%, because many sites lack these monitors.

In the data order, Step~1 gives the switching rates
\[
\hat\epsilon_{\rm on}=0.049\ (95\%\ \text{interval }0.017\text{--}0.131),\qquad \hat\epsilon_{\rm off}=0.026\ (0.008\text{--}0.082)\quad\text{per week}.
\]
A present link lasts about 38 weeks on average. Over the 120 orders, $\hat\epsilon_{\rm off}$ lies between 0.016 and 0.054, and the rate rule never selects the static graph.

\begin{figure}[!htbp]
\centering
\includegraphics[width=\linewidth]{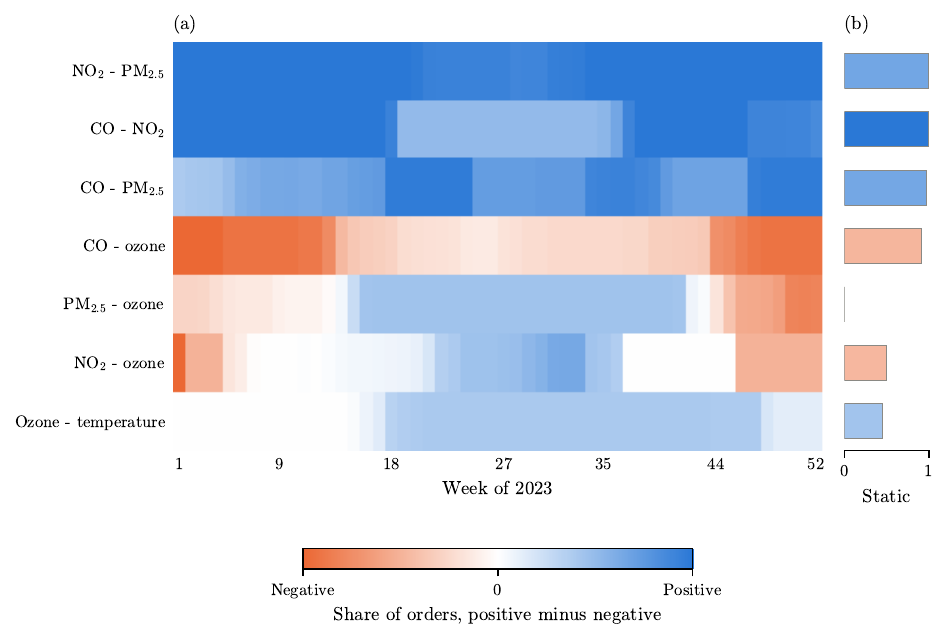}
\caption{Air quality, 2023. (a) Share of the 120 variable orders in which the Step-2 graph selects each link with a positive association, minus the share with a negative association, by week. Links selected in more than 20\% of the orders in some week. (b) Share of the orders in which the static graph of DBMW-0 selects each link. The bar colour gives the sign of the static association, darker for a stronger association.}\label{fig:aqsapp}
\end{figure}

Figure~\ref{fig:aqsapp}a shows, for each link and week, the share of the 120 orders that select the link with a positive association, minus the share with a negative association.
\begin{itemize}
\item \NOt--\PMt\ and CO--\NOt\ have a positive association throughout the year, in all orders. CO--\PMt\ joins them from February. The three pollutants share traffic and combustion sources.
\item CO--ozone has a negative association from November to March, in up to all orders, and is absent otherwise.
\item \NOt--ozone has a negative association in weeks 1--4 and 46--52, in up to all orders. In weeks 29--33 it has a positive association, in up to 63\% of the orders. In the data order, the associations $-b$ of the two signs are $-0.24$ and $+0.08$.
\item \PMt--ozone has a negative association in weeks 46--52, in up to 82\% of the orders, and a positive association in summer, in up to 44\% of them.
\end{itemize}
The negative winter links of ozone with \NOt\ and CO are the signature of titration by fresh emissions \citep{sillman1999}. The positive summer link of \NOt\ and ozone reflects photochemical production \citep{seinfeld2016}. Its weaker evidence matches the smaller summer association in Figure~\ref{fig:intro}b. The analysis thus recovers the known seasonal chemistry from the data alone.

The static graph of DBMW-0 gives a different picture (Figure~\ref{fig:aqsapp}b). It selects \NOt--ozone in half of the orders, with a weak negative association of $-0.06$, and \PMt--ozone in none of them. One static link averages the two regimes into a weak association of one sign, as anticipated in \S\ref{sec:intro}.

Table~\ref{tab:app} reports prediction. On whole withheld maps, DBMW and DBMW-X have errors of 0.187 and 0.188, against 0.334 for SPDE-AR1. The default SPDE mesh has 235 nodes for 288 sites, so nearby monitors share a node. Over the 120 orders, the median error is 0.196 for DBMW and 0.190 for DBMW-X. DBMW improves on the fit without a graph in 84\% of the orders, but on DBMW-X in only 28\% of them. For these continuous data with five variables, the joint refit thus sharpens the strengths but adds little to prediction. With withheld regions, DBMW has an error of 0.715, against 0.861 for the fit without a graph and 15.7 for SPDE-AR1. The large SPDE-AR1 error comes from \NOt. Its sites have strong, persistent levels that differ between nearby monitors. The AR(1) field represents these levels with a correlation near one and a large variance, and it can extrapolate far from the observed values into a withheld region. An added time-constant spatial field for the site levels lowers the error to 9.8, still more than ten times that of DBMW.

\begin{table}[t]
\centering
\caption{Prediction in the applications: loss on the withheld entries at the posterior mean of the expected response (squared error for air quality, deviance for surveillance). SPDE-AR1, site: SPDE-AR1 with an added time-constant spatial field for persistent site levels. Whole maps: one data set with 10\% of the maps withheld. Regions: one spatial region per variable withheld at all periods, means over three data sets. Orders: median over the 120 orderings of the air-quality variables. The best value in each row is in bold.}\label{tab:app}
\footnotesize
\begin{tabular}{llcccc}
\toprule
Data & Hold-out & SPDE-AR1 & SPDE-AR1, site & DBMW-X & DBMW \\
\midrule
Air quality & whole maps & 0.334 & 0.333 & 0.188 & \textbf{0.187} \\
 & whole maps, orders & -- & -- & \textbf{0.190} & 0.196 \\
 & regions & 15.717$^{a}$ & 9.792$^{a}$ & 0.720 & \textbf{0.715} \\
\addlinespace
Surveillance, fit A & whole maps & 7.23 & 7.57 & 7.28 & \textbf{6.33} \\
 & regions & 23.88 & 15.73 & 12.31 & \textbf{11.70} \\
\addlinespace
Surveillance, fit B & whole maps & 10.02 & 9.96 & 9.68 & \textbf{9.60} \\
 & regions & 9.91 & 10.64 & 5.53 & \textbf{5.48} \\
\bottomrule
\end{tabular}
\par\smallskip\parbox{0.9\linewidth}{\scriptsize $^{a}$Large errors for nitrogen dioxide in some of the withheld regions (Supplement~\Sref{sec:competitors}). Over the three data sets the mean squared error ranges from 2.9 to 25.9 for SPDE-AR1 and from 2.8 to 23.6 with the site field.}
\end{table}

\subsection{Infectious disease surveillance in Germany}\label{sec:survstat}
SurvStat@RKI 2.0 \citep{rki_survstat} reports the cases of notifiable diseases by district and week of notification. We retrieved the weekly counts of all 411 districts for 2016--2025. Seven diseases have complete national reporting over the whole period: influenza, norovirus, rotavirus, pertussis, varicella, campylobacteriosis and salmonellosis. Weekly counts are sparse, so we summed them into $T=130$ four-week periods. The counts total 4.4 million cases, with a mean per district and period from 2.5 for salmonellosis to 36.5 for influenza. The family is Poisson, and the fit uses 15 outer iterations.

The pandemic offers a natural experiment. Influenza and the enteric viruses follow strong winter seasons until 2020. Influenza then nearly vanished, with 1{,}129 notified cases in 2021 against 194{,}834 in 2020. Pertussis and rotavirus also fell sharply, and pandemic measures reduced most other notifiable diseases as well \citep{ullrich2021}. The analysis asks whether the dependence between diseases at the district level changed through this shock.

Two fits differ only in their offset. \emph{Fit A} uses log(population $\times$ weeks), so the fields carry the national seasonality and trend of each disease. \emph{Fit B} adds each disease's national log rate in each period, computed from the training entries only, so the fields describe district departures from the national curve. This offset plays the role of the endemic component of surveillance models \citep{held2005, paul2008}. It also contains the national drop of influenza, the situation of the vanishing-variable study of \S\ref{sec:simresults}.

In the data order, fit A gives $\hat\epsilon_{\rm off}=0.117$ (0.079--0.170) per period, and fit B gives $0.042$ (0.022--0.079). A present link lasts about nine periods in fit A and about 24 periods in fit B. The national curves thus absorb most of the switching.

\begin{figure}[!htbp]
\centering
\includegraphics[width=\linewidth]{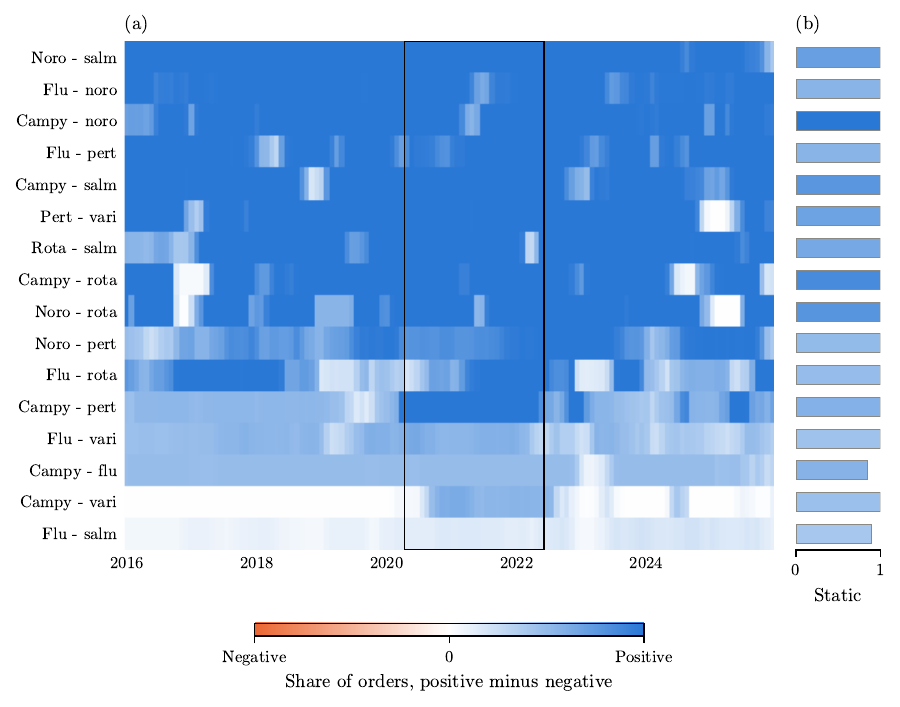}
\caption{Infectious disease surveillance, Germany, 2016--2025, fit B (national curve in the offset). (a) Share of the 120 variable orders in which the Step-2 graph selects each link with a positive association, minus the share with a negative association, by four-week period. The frame marks the pandemic period, May 2020 to June 2022. (b) Share of the orders in which the static graph of DBMW-0 selects each link, coloured as in Figure~\ref{fig:aqsapp}b. Abbreviations: flu influenza, noro norovirus, rota rotavirus, pert pertussis, vari varicella, campy campylobacteriosis, salm salmonellosis.}\label{fig:survapp}
\end{figure}

In fit A, the fields carry the national seasons, and links switch more often. Two links have a negative association: campylobacteriosis with influenza and salmonellosis with varicella (Supplement~\Sref{sec:apps}). These diseases peak in opposite seasons. Figure~\ref{fig:survapp} shows fit B. All 16 links that appear have a positive association, in every order and period, and the negative links of fit A disappear once the national curves enter the offset. Twelve links are selected in at least 64\% of the order-periods. Influenza keeps its links to norovirus, pertussis and rotavirus through the pandemic, with shares of 95\%, 97\% and 79\% in the pandemic periods against 98\%, 94\% and 65\% in the others. Two links of campylobacteriosis are more frequent during the pandemic: with pertussis, 99\% against 55\%, and with varicella, 48\% against 4\%. District-level dependence thus remains stable once the national curves enter the offset. One caution concerns reporting. Positive links between many diseases also arise from district differences in testing and reporting, a factor outside the model. Shared changes in reporting during the pandemic may explain the stronger campylobacteriosis links.

Table~\ref{tab:app} reports prediction. DBMW has the lowest deviance of all methods in both fits and both designs. In fit A, the joint refit lowers the deviance of DBMW-X by 13\% on whole maps. With withheld regions, DBMW lowers the deviance of the better SPDE-AR1 specification by 26\% in fit A and by 45\% in fit B. Supplement~\Sref{sec:apps} compares all estimators. In fit B, the national curves already carry the timing, and the fit without a graph predicts whole maps as well as any graph. In fit A, the static graph predicts whole maps best.

\section{Discussion}\label{sec:discussion}
DBMW learns a dependence graph whose links appear, vanish and change sign over time, from spatial data with Gaussian, count or binomial responses, with exact posterior inclusion probabilities given the fields. This exactness rests on one modelling choice: the temporal dynamics act on each field, and the graph acts on the temporal innovations, which keeps the normalizing constant free of the graph. In both applications, the learned links have clear scientific interpretations, and the graph improves the prediction of withheld data.

The method has limitations. The switching rates come from an on/off model, which slightly underestimates them when links change sign, and one pair of rates serves all links. A link changes sign only through absence, and short sign phases are hard to detect. Under a static truth, the rate rule chose the static graph in only 20--50\% of the data sets. The directed graph depends on the variable order, a dependence that averaging over orders removes at a cost linear in their number. Effects common to all variables, such as reporting intensity, produce positive links between many pairs.

Several extensions are natural. The bigraphical structure of space could carry over to other dimensions of the data. A graph Laplacian on the periods, in place of autoregression, would treat time as a second bigraphical dimension. Site-specific link states with a BMW prior over space would let the graph vary between regions as well as over time, as graphs over brain locations do in \citet{greenewald2017}. Other spatial operators, such as those of river or road networks and of fields on the sphere, fit the same construction as long as all variables share one eigenbasis. Link-specific, covariate-dependent and time-varying switching rates change only the factors of the transition matrix (Remark~\ref{rem:switch}). For overdispersed counts, a negative binomial likelihood with P\'olya--Gamma augmentation \citep{polson2013} would keep the E-step Gaussian. Finally, a joint search over variable orders and graphs would remove the need to average over orders.

\ifblind\else
\section*{Code availability}
The methods are implemented in the module \texttt{stbmw} of the \texttt{bmwspatial} package (\url{https://github.com/Ddey07/bmwspatial}). Scripts reproduce all numerical results.
\fi

\section*{Supplementary Material}
The Supplementary Material contains the proofs, details of the estimator, the choice of estimator, further simulation scenarios, the rate study, further hold-out designs, details of the competitors and further results of the applications.

\clearpage
\FloatBarrier
\setcounter{section}{0}\setcounter{equation}{0}\setcounter{table}{0}\setcounter{figure}{0}
\setcounter{algorithm}{0}\setcounter{lemma}{0}\setcounter{remark}{0}
\renewcommand{\thesection}{S\arabic{section}}\def\theHsection{S\arabic{section}}
\renewcommand{\theequation}{S\arabic{equation}}\def\theHequation{S\arabic{equation}}
\renewcommand{\thetable}{S\arabic{table}}\def\theHtable{S\arabic{table}}
\renewcommand{\thefigure}{S\arabic{figure}}\def\theHfigure{S\arabic{figure}}
\renewcommand{\thealgorithm}{S\arabic{algorithm}}\def\theHalgorithm{S\arabic{algorithm}}
\renewcommand{\thelemma}{S\arabic{lemma}}\def\theHlemma{S\arabic{lemma}}
\renewcommand{\theremark}{S\arabic{remark}}\def\theHremark{S\arabic{remark}}
\begin{center}
{\LARGE Supplementary Material}
\end{center}
\addcontentsline{toc}{section}{Supplementary Material}
\vspace{1ex}
\noindent Sections, equations, lemmas, remarks, tables, figures and algorithms with the prefix S belong to this supplement. Those without the prefix refer to the main text.

\FloatBarrier
\section{Proofs}\label{sec:proofs}
\begin{lemma}\label{lem:spec}
Every block of $\cL_t$ and of $Q_t$ is a function of $\Le$, $(\cL_t)_{ij}=\ell_{ij,t}(\Le)$ and $(Q_t)_{ij}=q_{ij,t}(\Le)$, with $\{\ell_{ij,t}(\lambda_h)\}_{ij}=L_{h,t}$ and $\{q_{ij,t}(\lambda_h)\}_{ij}=Q_{h,t}$.
\end{lemma}
\begin{proof}
$(\cL_t)_{ii}=d_i(\Le)$, and $(\cL_t)_{ij}=b_{ij}(t)I_n$ is a constant function of $\Le$. Also $(Q_t)_{ij}=\sum_k\tau_k^{-2}(\cL_t)_{ki}^\top(\cL_t)_{kj}=\sum_k\tau_k^{-2}\ell_{ki,t}(\Le)\ell_{kj,t}(\Le)$. Functions of one symmetric matrix commute and multiply pointwise on its spectrum. Evaluation at $\lambda_h$ gives $(\tilde L_{h,t}^\top\tilde L_{h,t})_{ij}$.
\end{proof}

\begin{proof}[Proof of Theorem~\Mref{thm:modes}]
We write $\mathcal U=I_p\otimes U$. By Lemma~\ref{lem:spec} and Assumption~\Mref{ass:R}, each $n\times n$ block of $\mathcal U^\top Q_t\,\mathcal U$ and of $\mathcal U^\top R\,\mathcal U$ is diagonal, with $h$th entry $q_{ij,t}(\lambda_h)$ and $r_i(\lambda_h)\delta_{ij}$ respectively. After permuting coordinates from (variable, mode) to (mode, variable) order, $\mathcal U^\top Q_t\mathcal U=\blkdiag_h(Q_{h,t})$ and $\mathcal U^\top R\,\mathcal U=\blkdiag_h(\Phi_h)$. The $h$th block of $\mathcal U^\top$ applied to $w_t=Rw_{t-1}+\eta_t$ gives $\omega_{h,t}=\Phi_h\omega_{h,t-1}+\xi_{h,t}$ with $\xi_{h,t}=(I_p\otimes u_h^\top)\eta_t$. Since $\mathrm{cov}(\mathcal U^\top\eta_t)=\mathcal U^\top Q_t^{-1}\mathcal U$ is block diagonal over modes with blocks $Q_{h,t}^{-1}$, and the $\eta_t$ are independent over $t$, the $\xi_{h,t}$ are Gaussian and independent over $(h,t)$.
\end{proof}

\begin{proof}[Proof of Theorem~\Mref{thm:logdet}]
The map $w_{1:T}\mapsto\eta_{1:T}=Gw_{1:T}$ is linear, and $G$ is block lower triangular with identity diagonal blocks, so $\det G=1$ for every $R$. The density of $w_{1:T}$ is therefore $\prod_tp_{\eta_t}\{(Gw_{1:T})_t\}$, and $Q_{w(1:T)}=G^\top\blkdiag(Q_1,\dots,Q_T)G$. Hence $\log|Q_{w(1:T)}|=\sum_t\log|Q_t|$. Each $\cL_t$ is block lower triangular, so $\log|\det\cL_t|=\sum_j\log|\det d_j(\Le)|=\sum_j(\alpha_j/2)\sum_h\log(\kappa_j^2+\lambda_h)$, which does not depend on $t$. Finally $\log|Q_t|=2\log|\det\cL_t|-2n\sum_j\log\tau_j$.
\end{proof}

\begin{remark}[Initial condition]\label{rem:stationary}
Theorem~\Mref{thm:logdet} uses the initial condition $w_0=0$, under which $w_1=\eta_1$. Under the stationary initial law with $R=\phi I$ and a static graph, $\Sigma_\infty=Q^{-1}/(1-\phi^2)$ solves the Lyapunov equation, and replacing the first block $Q$ by $(1-\phi^2)Q$ adds the constant $np\log(1-\phi^2)$. With variable-specific persistence, the stationary covariance $\sum_{m\ge0}R^mQ^{-1}R^m$ involves the coupling coefficients, and the graph-free property is lost. We therefore use the conditional initial condition throughout.
\end{remark}

\begin{proof}[Proof of Theorem~\Mref{thm:markov}]
Expanding $G^\top\blkdiag(Q_1,\dots,Q_T)G$ gives three kinds of time block: the diagonal time blocks $Q_t+RQ_{t+1}R$ for $t<T$ and $Q_T$ for $t=T$, the time blocks at lag one, $-Q_{t+1}R$ in position $(t+1,t)$ with their transposes, and zero blocks otherwise. Since $R$ is block diagonal over variables with blocks $r_i(\Le)$, the variable block $(i,j)$ of the diagonal time block is
\[
q_{ij,t}(\Le)+r_i(\Le)q_{ij,t+1}(\Le)r_j(\Le).
\]
By \citet[Theorem~1]{dey2026bmw}, $q_{ij,t}\equiv0$ if and only if $i,j$ are non-adjacent in $\Mor(\GV_t)$. If exactly one of $q_{ij,t}$ and $q_{ij,t+1}$ is identically zero, the block is nonzero, because $r_ir_j$ is nonzero. If both are identically zero, the block is zero. If neither is, the block vanishes only if $q_{ij,t}(\lambda_h)+r_i(\lambda_h)r_j(\lambda_h)q_{ij,t+1}(\lambda_h)=0$ for every $h$. For fixed $h$, the left-hand side is a polynomial in the coupling coefficients $\{b_{kl}(t),b_{kl}(t+1)\}$. It is not identically zero, since setting the coefficients at time $t+1$ to zero leaves $q_{ij,t}(\lambda_h)$, which is nonzero for generic coefficients. The zero set of a non-trivial polynomial has Lebesgue measure zero. For a static graph, $q_{ij,t}=q_{ij,t+1}$ and the block $(1+r_ir_j)q_{ij}$ vanishes exactly when $q_{ij}$ does, because $1+r_ir_j>0$. The variable block $(i,j)$ of a lag-one block is $-q_{ij,t+1}(\Le)r_j(\Le)$. Because $r_j(\lambda)\neq0$, it vanishes if and only if $q_{ij,t+1}\equiv0$. For $i=j$, $q_{ii,t+1}(\lambda)\ge\tau_i^{-2}d_i(\lambda)^2>0$. For a static graph the two cases combine into the strong product of $\Mor(\GV)$ with the path on $T$ vertices.
\end{proof}

\begin{proof}[Proof of Corollary~\Mref{cor:kalman}]
The rotation $\mathcal U^\top$ maps white noise to white noise. By Theorem~\Mref{thm:modes}, the rotated data are independent over $h$, and each mode follows a linear Gaussian state-space model. Its likelihood is given exactly by the Kalman filter.
\end{proof}

\begin{proof}[Proof of Lemma~\Mref{lem:perturb}]
$u$ is linear in independent standard normal vectors. Its covariance has $\theta$-block $\cA^\top\cA+X^\top WX=Q_\theta+H$, cross-block $X^\top W\mathbf 1=H_{\theta a}$ and $a$-block $\mathbf 1^\top W\mathbf 1+\gamma_aI=H_{aa}+\gamma_aI$. Hence $\mathrm{cov}(u)=P$ and $\mathrm{cov}(P^{-1}u)=P^{-1}$.
\end{proof}

\begin{proof}[Proof of Proposition~\Mref{prop:fb}]
By \eqref{eq:Qsplit}, $\mathcal Q$ is a sum over $(i,t)$ of terms that each depend on $\mathcal B$ only through the row $b_{i\cdot}(t)$. With the ordering fixed, every assignment of parent sets gives an acyclic graph. Given $\beta$, the prior of \S\Mref{sec:graphprior} is a product over links of independent augmented chains and independent deviations $\delta_{ij}(t)$. It therefore factorizes over children, and given the states it factorizes over time. For child $i$ with sign configuration $c$ at time $t$, the $(i,t)$ term is $c_{i,t}+\mathsf g^\top b-\tfrac12b^\top\mathsf Gb$ with $b=\mu_c+\delta$ on the support of $c$. Completion of the square and integration against $N(\delta;0,s_\delta^2I)$ gives $e^{c_{i,t}}\exp\{e_{i,t}(c;\beta)\}$. The joint augmented state of the $|\mathcal C_i|$ links of child $i$ is a Markov chain on $8^{|\mathcal C_i|}$ states, with the Kronecker product of the link transition matrices as transition matrix. The Occam factors of \S\Mref{sec:graphstep} multiply single transitions of single links, so they keep this product form. Given the chain, the emissions are independent over $t$ and depend on the joint state only through the current signs. The posterior is therefore that of a hidden Markov model, and the forward--backward algorithm computes its marginals exactly. A product of a vector on $8^{K}$ states with a Kronecker product of $K$ matrices of size $8\times8$ costs $K$ mode products of $O(8^{K+1})$ operations, and each of the $3^K$ emissions costs $O(K^3)$.
\end{proof}

With $\epsilon_{\rm off}=0$, the parent set $\mathcal S_i$ and the coefficients $b_{i\mathcal S}(t)=\beta_{i\mathcal S}$ do not change over time. A static link with a random sign and a half-normal strength has a $N(0,s^2)$ coefficient, so the sum over $t$ of the $(i,t)$ terms is $\sum_tc_{i,t}+\mathsf g_{i,\mathcal S}^\top\beta-\tfrac12\beta^\top\mathsf G_{i,\mathcal S}\beta$ with $\mathsf G_{i,\mathcal S}=\sum_t\mathsf G_{i,t,\mathcal S}$ and $\mathsf g_{i,\mathcal S}=\sum_t\mathsf g_{i,t,\mathcal S}$. Integration against $N(0,s^2I)$ gives the closed-form posterior
\[
\Pr(\mathcal S_i=\mathcal S)\propto\pi^{|\mathcal S|}(1-\pi)^{|\mathcal C_i|-|\mathcal S|}\exp\Big\{-\tfrac12\log|I+s^2\mathsf G_{i,\mathcal S}|+\tfrac12\mathsf g_{i,\mathcal S}^\top(\mathsf G_{i,\mathcal S}+s^{-2}I)^{-1}\mathsf g_{i,\mathcal S}\Big\},
\]
computed by enumeration of the parent sets. This is the graph step of DBMW-0.

\subsection{The graph as a property of the process}\label{sec:markovex}
In a Gaussian process the conditional independences are the zero blocks of the precision matrix. Theorem~\Mref{thm:markov} lists every nonzero block of $Q_{w(1:T)}$, the precision matrix of all fields at all times, and each condition depends only on the graphs $\GV_1,\dots,\GV_T$. The graph sequence therefore determines the conditional-independence structure of the whole space-time field. Three kinds of dependence appear.
\begin{enumerate}
\item The fields of two variables in the same period $t$ are conditionally dependent when the variables are adjacent in the moral graph at $t$ or at $t+1$. The period $t+1$ enters because $w_t$ also appears in the innovation $\xi_{t+1}=w_{t+1}-Rw_t$.
\item The field of variable $i$ at $t+1$ and that of variable $j$ at $t$ are conditionally dependent when $i$ and $j$ are adjacent in the moral graph at $t+1$.
\item The field of each variable depends on its own field in the neighbouring periods, through the persistence.
\end{enumerate}
An example has $p=3$ and a single link $3\leftarrow1$, present only at $t_0$. Here $w_{1t}$ and $w_{3t}$ are conditionally dependent for $t=t_0-1$ and $t=t_0$, and so are $w_{1,t_0-1}$ and $w_{3,t_0}$, and $w_{3,t_0-1}$ and $w_{1,t_0}$. At all other times the fields of variables 1 and 3 are conditionally independent given all other fields. Variable 2 has no link, so its field is conditionally independent of the other two at all times. A link that switches on and off thus changes the conditional independences of the fields themselves, exactly and only around the periods of its presence. This differs from a graph estimated from fitted residuals, as by the graphical lasso on innovations, which describes the estimate rather than the process that generated the data.

\FloatBarrier
\section{Details of the estimator}\label{sec:details}
\subsection{The augmented link chain and the strengths}\label{sec:chain}
Each state of the augmented chain is a pair (current state, signs used so far). The eight states are
\[
(0,\emptyset),\ (0,\{+\}),\ (0,\{-\}),\ (0,\{+,-\}),\ (+,\{+\}),\ (+,\{+,-\}),\ (-,\{-\}),\ (-,\{+,-\}).
\]
Table~\ref{tab:chain} lists the transitions. From a present state, the link leaves with probability $\epsilon_{\rm off}$ and keeps its set of used signs. From an absent state, it enters with probability $\epsilon_{\rm first}$ if no sign has been used and $\epsilon_{\rm on}$ otherwise, and it chooses either sign with probability $1/2$. An entry with a sign not used before carries the Occam factor $e^{o_{ij}}$ of \S\Mref{sec:graphstep}. The initial law puts $1-\pi$ on $(0,\emptyset)$ and $\pi e^{o_{ij}}/2$ on each of $(+,\{+\})$ and $(-,\{-\})$. The memory of used signs serves only the Occam factor and the separation of never-present candidates. The law of the current states $\gamma_{ij}(\cdot)$ is the chain \eqref{eq:edgechain}.

\begin{table}[!htbp]
\centering
\caption{Transitions of the augmented link chain out of each state, other than staying. $\epsilon_{\rm first}=\epsilon_{\rm off}\pi/(1-\pi)$. Occ.: the transition carries the Occam factor $e^{o_{ij}}$.}\label{tab:chain}
\footnotesize
\begin{tabular}{lll}
\toprule
From & To, with probability & Occ. \\
\midrule
$(0,\emptyset)$ & $(+,\{+\})$ and $(-,\{-\})$, each $\epsilon_{\rm first}/2$ & yes \\
$(0,\{+\})$ & $(+,\{+\})$, $\epsilon_{\rm on}/2$, and $(-,\{+,-\})$, $\epsilon_{\rm on}/2$ & the second \\
$(0,\{-\})$ & $(-,\{-\})$, $\epsilon_{\rm on}/2$, and $(+,\{+,-\})$, $\epsilon_{\rm on}/2$ & the second \\
$(0,\{+,-\})$ & $(+,\{+,-\})$ and $(-,\{+,-\})$, each $\epsilon_{\rm on}/2$ & no \\
$(\pm,U)$ & $(0,U)$, $\epsilon_{\rm off}$ & no \\
\bottomrule
\end{tabular}
\end{table}

For child $i$, collect the strengths in $\theta_i=(-\beta^+_{i\cdot},\beta^-_{i\cdot})\in\mathbb R^{2|\mathcal C_i|}$, so that $b_{i\cdot}(t)=Z_t\theta_i$ with indicator matrix $Z_t=(\mathrm{diag}\,1\{c_t=+1\},\mathrm{diag}\,1\{c_t=-1\})$. In the limit $s_\delta\to0$, the expected complete-data objective is
\[
\sum_t\E\big\{\mathsf g_{i,t}^\top Z_t\theta_i-\tfrac12\theta_i^\top Z_t^\top\mathsf G_{i,t}Z_t\theta_i\big\}-\frac{\|\theta_i\|^2}{2s^2}=c^\top\theta_i-\tfrac12\theta_i^\top A\theta_i,
\]
with $A=\sum_t\E(Z_t^\top\mathsf G_{i,t}Z_t)+s^{-2}I$ and $c=\sum_t\E(Z_t)^\top\mathsf g_{i,t}$, where the expectations are over the posterior of the sign configurations. With the Cholesky factor $A=R^\top R$, the update solves the bounded least-squares problem $\min\|R\theta_i-R^{-\top}c\|^2$ with the first $|\mathcal C_i|$ entries nonpositive and the others nonnegative. The starting values come from multiple regressions at each period. For each link and sign, the start is the precision-weighted mean of the per-period coefficients of that sign with $|z|>1.5$, and $\pm0.05$ when there are none. The number of alternations is fixed at ten. In one data set each of S1P and S1PF, the largest change of a strength between ten and forty alternations is 0.0012 and 0.0022, or 0.7\% and 1.8\% of the largest strength, and the largest change of an inclusion probability is 0.0015 and 0.0003. After ten alternations, a sign whose strength sits at its bound is removed from the state space of its link, because a present link with zero strength would describe absence plus a random deviation.

Link-specific rates with $\epsilon_{{\rm off},ij}\sim\mathrm{Beta}(a_\epsilon,b_\epsilon)$ give the Baum--Welch update $(N^{ij}_{10}+a_\epsilon-1)/(N^{ij}_{10}+N^{ij}_{11}+a_\epsilon+b_\epsilon-2)$, with $N^{ij}_{10}$ and $N^{ij}_{11}$ the expected numbers of transitions from present to absent and from present to present. Covariate-dependent rates, $\mathrm{logit}\,\epsilon_{{\rm off},ij}=x_{ij}^\top\vartheta$, give a weighted logistic regression on the expected transition counts. Rates that vary with $t$ give time-inhomogeneous chains. In each case only the factors of the Kronecker product change, and the forward--backward recursions stay exact at the same cost.

\subsection{The switching-rate step}\label{sec:ratestep}
\begin{enumerate}
\item The fit without a graph, with $\mathcal B=0$, gives $S'=20$ draws from $q$. For each $(i,t,h)$, $m_{i,t,h}$ and $v_{i,t,h}$ denote the mean and variance of the drawn innovations $\xi_{h,t,i}$. Their prior variance under the fit without a graph is $\tau_i^2/d_i(\lambda_h)^2$. Its removal gives a pseudo-observation of the innovation with error variance $R_{i,t,h}$,
\[
R_{i,t,h}^{-1}=v_{i,t,h}^{-1}-d_i(\lambda_h)^2/\tau_i^2,\qquad y^*_{i,t,h}=(R_{i,t,h}/v_{i,t,h})\,m_{i,t,h}.
\]
Modes without information from the data, with $R_{i,t,h}=\infty$, drop out of the regression.
\item For each child $i$ and time $t$, a weighted regression of $d_i(\lambda_h)y^*_{i,t,h}$ on $\{m_{j,t,h}\}_{j\in\mathcal C^\epsilon_i}$ over the modes $h$, with weights $w_{i,t,h}=\tau_i^2/\{\tau_i^2+d_i(\lambda_h)^2R_{i,t,h}\}$, gives a series $\hat b_{ij}(t)$ with working standard errors $s_{ij}(t)$. The candidates $\mathcal C^\epsilon_i$ are the earlier variables with $|z|>3$ in a regression pooled over periods and modes.
\item The working standard error $s_{ij}(t)$ is proportional to the amplitude $\hat\tau_i$ of the fit without a graph, which absorbs the contribution of the parents of $i$. It also ignores the error in the parent's innovation, which enters the regression multiplied by $b_{ij}(t)$ and so acts only while the link is present. The first distortion is specific to the child, and the second to the state of the link. With the emissions \eqref{eq:calemission}, a two-state chain with $\Pr(0\to1)=\epsilon_{\rm on}$, $\Pr(1\to0)=\epsilon_{\rm off}$ and a free initial probability $p_1$, and a share $\varrho$ of candidates that are links, the log-likelihood of $\vartheta=(\epsilon_{\rm on},\epsilon_{\rm off},\varrho,p_1,c_1,c_{0,1},\dots,c_{0,p})$ is
\begin{multline*}
\ell(\vartheta)=\sum_{(i,j)\in\mathcal C^\epsilon}\log\Big[(1-\varrho)\prod_tN\{\hat b_{ij}(t);0,c_{0,i}^2s_{ij}(t)^2\}\\
+\varrho\int\sum_{\gamma(\cdot)}\Pr\{\gamma(\cdot)\mid\epsilon_{\rm on},\epsilon_{\rm off},p_1\}\prod_tN\big[\hat b_{ij}(t);\gamma(t)\beta,\{c_{0,i}^{1-\gamma(t)}c_1^{\gamma(t)}\}^2s_{ij}(t)^2\big]\,N(\beta;0,s^2)\,d\beta\Big].
\end{multline*}
It is computed exactly by forward recursions, with $\beta$ integrated on a grid of 201 points. An EM algorithm maximizes it. Its E-step is the forward--backward recursion for each candidate. In its M-step, $\epsilon_{\rm on}$ and $\epsilon_{\rm off}$ are Baum--Welch ratios of expected transition counts \citep{baum1970}, $\varrho$ and $p_1$ are posterior means, and $c_{0,i}^2$ and $c_1^2$ are weighted means of the squared standardized coefficients in the absent and present states. The scale factors are bounded below by 0.05. The mixture keeps candidates that are never links from pulling $\epsilon_{\rm on}$ towards zero.
\item The 95\% intervals are Wald intervals on the logit scale, from the observed information at the maximum, computed numerically over the logits of $\epsilon_{\rm on}$, $\epsilon_{\rm off}$, $\varrho$ and $p_1$ and the logarithms of $c_1$ and $\bar c_0$. Here $\bar c_0$ rescales all $c_{0,i}$ jointly, and their ratios are held at the estimates.
\end{enumerate}
When $\epsilon_{\rm on}$ is not identified, Step~2 uses $\epsilon_{\rm first}$ for re-entry. Step~2 multiplies the scale $\tau_i$ of each child by $\max(1,\hat c_{0,i})$. A two-sided factor, which also narrows the scale when $\hat c_{0,i}<1$, multiplied the false links per period by five to eight in the simulations, where $\hat c_{0,i}$ lay between 0.45 and 0.9.

\subsection{Parameter updates}\label{sec:params}
Under scalar persistence, the update of $\phi$ solves the $p\times p$ system
\[
\Big\{\E_q\sum_{h,t\ge2}\diag(\omega_{h,t-1})\,Q_{h,t}\,\diag(\omega_{h,t-1})\Big\}\phi=\E_q\sum_{h,t\ge2}\diag(\omega_{h,t-1})\,Q_{h,t}\,\omega_{h,t}.
\]
For each $i$, the amplitude has the closed form $\hat\tau_i^2(\kappa_i)=(nT)^{-1}\E_q\sum_{h,t}\{d_i(\lambda_h)\xi_{h,t,i}+\sum_jb_{ij}(t)\xi_{h,t,j}\}^2$. The profile $T\sum_h\log d_i(\lambda_h)-\tfrac{nT}{2}\log\hat\tau_i^2(\kappa_i)$ is maximized over $\log\kappa_i$ by a bounded one-dimensional search, and over $\alpha_i$ on the grid $\{1.2,1.4,\dots,3.6\}$. The fit without a graph uses $\alpha_i=2$, because $\alpha_i$ is weakly identified from counts. In the link-aware rounds of Step~1, pooled least squares over draws, times and modes give the coefficient of each rate candidate,
\[
\hat\beta_{i\mathcal C_i}=-\Big(\sum_t\mathsf G_{i,t,\mathcal C_i}\Big)^{-1}\sum_t\mathsf c_{i,t,\mathcal C_i},\qquad \mathsf c_{i,t,\mathcal C_i}=\tau_i^{-2}\E_q\sum_hd_i(\lambda_h)\xi_{h,t,i}\xi_{h,t,\mathcal C_i},
\]
and the range, amplitude and persistence updates use these coefficients. For the Gaussian family, $\sigma^2$ is the mean squared residual over paired draws of the fields and the intercepts, and every later fit starts from it.

\subsection{Matrix-free computation}\label{sec:matfree}
A product with $Q_\theta$ costs two batched products with the $p\times p$ matrices $\tilde L_{h,t}$, at $O(Tnp^2)$, and a product with $H$ costs two rotations per variable--time pair. Systems with $P$ are solved by preconditioned conjugate gradients, to a relative residual of $10^{-6}$ and at most 2000 iterations.
\begin{itemize}
\item The preconditioner is block diagonal over variable--time pairs. The block of $(i,t)$ is the exact data block $U^\top W_{it}U$ plus the diagonal of $Q_\theta$ on the modes of $(i,t)$. The $Tp$ blocks of size $n\times n$ are inverted at cost $O(Tpn^3)$. They are rebuilt only when the median of $|\log W_{it}/W^{\rm last}_{it}|$ exceeds 0.15.
\item The likelihood does not change when a constant is added to the field of variable $i$ at every time and subtracted from $a_i$, and the prior penalizes the constant field only weakly. These $p$ directions make $P$ ill-conditioned, and without deflation conjugate gradients needed several hundred to over two thousand iterations. The $2p$ directions spanned by the constant fields and the intercepts are solved exactly, through a $2p\times2p$ system \citep{tang2009deflation}, and conjugate gradients runs on the complement.
\end{itemize}

\begin{algorithm}[!htbp]
\caption{Laplace--Monte Carlo EM for the DBMW model}\label{alg:mcem}
\begin{algorithmic}[1]
\Require responses, operator $\Le$ with eigenpairs, candidate links $\mathcal C$, switching rates (dynamic graph), draws $S$, outer iterations $O$, starting graph and parameters
\For{$o=1,\dots,O$}
  \State Newton with preconditioned conjugate gradients for $(\hat\theta,\hat a)$, warm-started
  \State draw $\omega^{(1:S)}\sim q$ (Lemma~\Mref{lem:perturb}) and form the innovations
  \State for each child: statistics $\mathsf G_{i,t},\mathsf g_{i,t}$, signed forward--backward (Proposition~\Mref{prop:fb}) and strengths, update of $B_t$ from the median-probability graph
  \State persistence update, then range and amplitude updates, then $\sigma^2$ for the Gaussian family
\EndFor
\State \Return inclusion probabilities and signs, $\mathcal B$, $\psi$, the posterior mode and draws of the fields
\end{algorithmic}
\end{algorithm}

\begin{remark}
The graph posterior of Proposition~\Mref{prop:fb} conditions on the E-step distribution, a variational approximation. It can therefore understate the uncertainty about the graph. A fully Bayesian alternative uses delayed-acceptance Metropolis--Hastings: proposals are screened with $\mathcal Q$ and accepted with the Laplace marginal likelihood, where the change in $\log|P|$ is estimated by stochastic Lanczos quadrature \citep{ubaru2017}.
\end{remark}

\FloatBarrier
\section{Choice of estimator}\label{sec:choice}
An alternative to DBMW runs Algorithm~\ref{alg:mcem} directly from the fit without a graph, first with the static graph and then with the signed switching graph from the static fit. We call this the joint EM fit (Joint). Table~\ref{tab:Sest} compares it with DBMW on the same data. Joint reaches a higher average precision than DBMW in S1G and under a static truth, and a lower or equal one in the other scenarios. Its inclusion probabilities are overconfident. At most 3\% of its probabilities lie between 0.1 and 0.9, against 6--18\% for DBMW, and of its probabilities above 0.9 a smaller share are true links in every scenario with a switching truth. The cause is the feedback of the prior coupling of an included link into its own evidence. DBMW avoids it by computing the graph posterior once, from a fit without a graph. Joint predicts whole maps slightly better than DBMW in S1G and M1 (Table~\ref{tab:Spred}). Table~\ref{tab:Stime} gives the run times by step. Joint costs about as much as DBMW.

\begin{table}[t]
\centering
\caption{Choice of estimator: DBMW against the joint EM fit of the switching graph (Joint) on the same data. AP and ECE as in Tables~\ref{tab:graph} and \ref{tab:unc}. Confident: share of (link, period) pairs with inclusion probability above 0.9 that are true links. Middle: share of probabilities between 0.1 and 0.9. Minutes: timing study, means over three data sets.}\label{tab:Sest}
\footnotesize
\begin{tabular}{lcccccccccc}
\toprule
& \multicolumn{2}{c}{AP} & \multicolumn{2}{c}{ECE} & \multicolumn{2}{c}{Confident} & \multicolumn{2}{c}{Middle} & \multicolumn{2}{c}{Minutes} \\
\cmidrule(lr){2-3}\cmidrule(lr){4-5}\cmidrule(lr){6-7}\cmidrule(lr){8-9}\cmidrule(lr){10-11}
Scenario & DBMW & Joint & DBMW & Joint & DBMW & Joint & DBMW & Joint & DBMW & Joint \\
\midrule
S1G & 0.67 & 0.75 & 0.039 & 0.035 & 0.94 & 0.90 & 0.08 & 0.03 & 11.2 & 13.6 \\
S1P & 0.68 & 0.64 & 0.039 & 0.054 & 0.93 & 0.87 & 0.10 & 0.02 & 20.9 & 17.9 \\
S1N & 0.63 & 0.60 & 0.042 & 0.057 & 0.95 & 0.84 & 0.09 & 0.02 & 21.8 & 21.0 \\
S1PF & 0.71 & 0.63 & 0.027 & 0.044 & 0.98 & 0.88 & 0.06 & 0.01 & 37.3 & 33.6 \\
S1GF & 0.72 & 0.70 & 0.020 & 0.029 & 0.97 & 0.94 & 0.06 & 0.01 & 24.9 & 21.8 \\
S1B & 0.43 & 0.24 & 0.057 & 0.087 & 0.88 & 0.81 & 0.09 & 0.00 & 18.4 & 22.3 \\
M1 & 0.53 & 0.51 & 0.042 & 0.061 & 0.80 & 0.68 & 0.18 & 0.03 & -- & -- \\
M2 & 0.61 & 0.61 & 0.038 & 0.049 & 0.96 & 0.84 & 0.10 & 0.02 & -- & -- \\
M3 & 0.63 & 0.61 & 0.050 & 0.060 & 0.91 & 0.83 & 0.08 & 0.01 & -- & -- \\
SCP & -- & -- & 0.445 & 0.446 & 0.00 & 0.00 & 0.27 & 0.02 & -- & -- \\
S2G & 0.86 & 0.91 & 0.078 & 0.031 & 0.99 & 1.00 & 0.08 & 0.00 & -- & -- \\
S2P & 0.76 & 0.84 & 0.108 & 0.058 & 0.93 & 0.96 & 0.08 & 0.00 & -- & -- \\
\bottomrule
\end{tabular}
\end{table}
\begin{table}[t]
\centering
\caption{Run time in minutes per data set on two cores, by step, means over three data sets (one data set for the applications). Joint EM: the joint fits of the static and the switching graph, including step 1.}\label{tab:Stime}
\footnotesize
\begin{tabular}{lcccccc}
\toprule
Data & Step 1 & Step 2 & Step 3 & DBMW-X & DBMW & Joint EM \\
\midrule
S1G & 4.8 & 1.2 & 5.1 & 6.1 & 11.2 & 13.6 \\
S1P & 9.4 & 1.9 & 9.6 & 11.3 & 20.9 & 17.9 \\
S1N & 10.1 & 2.0 & 9.6 & 12.1 & 21.8 & 21.0 \\
S1B & 7.8 & 1.8 & 8.8 & 9.6 & 18.4 & 22.3 \\
S1GF & 11.1 & 2.5 & 11.2 & 13.6 & 24.9 & 21.8 \\
S1PF & 18.5 & 3.7 & 15.1 & 22.2 & 37.3 & 33.6 \\
Air quality & 14.0 & 3.8 & 20.8 & 17.8 & 38.7 & 34.1 \\
Surveillance, fit A & 42.9 & 6.4 & 26.0 & 49.2 & 75.2 & 51.1 \\
Surveillance, fit B & 40.3 & 6.3 & 27.3 & 46.6 & 73.9 & 47.8 \\
\bottomrule
\end{tabular}
\end{table}

\FloatBarrier
\section{Further simulation scenarios}\label{sec:other}
Table~\ref{tab:Ssettings} lists the further scenarios. All share the design of \S\Mref{sec:simdesign} unless stated otherwise. S1PFS repeats S1PF with doubled strengths. The graph baselines were run on the scenarios with a switching or static graph and fixed or changing signs. Tables~\ref{tab:Sgraph}--\ref{tab:Sphase} report graph recovery, the baselines on the raw responses, prediction with all estimators and both competitors, the uncertainty measures and the detection of sign phases.

\begin{table}[!htbp]
\centering
\caption{Further simulation scenarios. All have $n=200$ locations and $p=20$ variables, with $T=26$ periods unless stated otherwise.}\label{tab:Ssettings}
\footnotesize
\begin{tabular}{lll}
\toprule
Scenario & Responses & Links over time \\
\midrule
S1GF & Gaussian & $T=52$, $\epsilon=0.05$, new random sign at each entry \\
S1PFS & Poisson & as S1PF, strengths doubled \\
S1B & binary, logit link & $\epsilon=0.1$, fixed sign \\
M1 & negative binomial & as S1P, size 2 \\
M2 & Poisson & as S1P, $\mathrm{logit}\,\epsilon_{ij}\sim N(\mathrm{logit}\,0.1,1)$ \\
M3 & Poisson & as S1P, linear ramps over four periods around each switch \\
SCP & Poisson & no graph, a shared AR(1) Mat\'ern component, loadings $\pm\mathrm{Un}(0.5,1)\,\mathrm{sd}(w)$ \\
S2G, S2P & Gaussian, Poisson & static graph, every true link present at all times \\
\bottomrule
\end{tabular}
\end{table}

With Gaussian sign changes (S1GF), DBMW keeps the highest average precision, 0.72, and GL falls to 0.26. Table~\ref{tab:Sphase} splits the sign phases of \S\Mref{sec:simresults} by the type of link. At every phase length, a phase of a link whose truth has both signs is found less often than a phase of a link with one sign. Doubled strengths (S1PFS) raise detection for both types of link but keep the gap, and the joint refit of Step~3 finds both phases for 35\% of the links that change sign in S1PFS. Binary responses (S1B) carry little information about the graph for every method. Overdispersion (M1) raises the realized false discovery rate to 0.32. Link-specific rates (M2) and gradual switches (M3) change the results little. Under a static truth (S2G, S2P), GL matches DBMW in average precision but selects 42--45 false links per period. The rate rule chooses the static graph in 50\% and 20\% of these data sets, and the joint static fit (Joint-0) then predicts better than DBMW. Under the shared component (SCP), the truth has no graph. The shared component makes the fields of all variables dependent, and every graph estimator represents this dependence by links, about 37 per period for DBMW and Joint. The multivariate CAR model diverges on whole withheld count maps (S1P, M1, SCP), for the plug-in, the posterior-mean and the median prediction alike.

\begin{table}[t]
\centering
\caption{Graph recovery in the other scenarios, as in Table~\ref{tab:graph}, with the joint EM fit of the switching graph (Joint).}\label{tab:Sgraph}
\footnotesize
\begin{tabular}{lcccccccccccc}
\toprule
& \multicolumn{2}{c}{DBMW} & \multicolumn{2}{c}{GL} & \multicolumn{2}{c}{GL-W} & \multicolumn{2}{c}{TVGL} & \multicolumn{2}{c}{PLN} & \multicolumn{2}{c}{Joint} \\
\cmidrule(lr){2-3}\cmidrule(lr){4-5}\cmidrule(lr){6-7}\cmidrule(lr){8-9}\cmidrule(lr){10-11}\cmidrule(lr){12-13}
Scenario & AP & FP & AP & FP & AP & FP & AP & FP & AP & FP & AP & FP \\
\midrule
S1GF & \textbf{0.72} & 0.7 & 0.26 & 4.5 & 0.43 & 18.4 & 0.31 & 172.2 & -- & -- & 0.70 & 0.5 \\
S1B & 0.43 & 0.8 & \textbf{0.44} & 21.1 & 0.25 & 20.1 & 0.16 & 175.0 & -- & -- & 0.24 & 0.3 \\
M1 & \textbf{0.53} & 3.1 & -- & -- & -- & -- & -- & -- & -- & -- & 0.51 & 2.7 \\
M2 & \textbf{0.61} & 1.3 & -- & -- & -- & -- & -- & -- & -- & -- & \textbf{0.61} & 1.1 \\
M3 & \textbf{0.63} & 1.1 & -- & -- & -- & -- & -- & -- & -- & -- & 0.61 & 1.3 \\
SCP & -- & 37.1 & -- & -- & -- & -- & -- & -- & -- & -- & -- & 38.0 \\
S2G & 0.86 & 0.4 & 0.85 & 42.1 & 0.58 & 39.5 & 0.43 & 162.1 & -- & -- & \textbf{0.91} & 0.0 \\
S2P & 0.76 & 0.6 & 0.79 & 45.0 & 0.50 & 39.5 & 0.36 & 162.5 & 0.54 & 17.2 & \textbf{0.84} & 0.5 \\
S1PFS & \textbf{0.70} & 1.0 & -- & -- & -- & -- & -- & -- & -- & -- & 0.68 & 1.4 \\
\bottomrule
\end{tabular}
\end{table}
\begin{table}[t]
\centering
\caption{Graph recovery of the baselines applied to the standardized log counts or responses instead of the innovations: AP (FP), means over ten data sets.}\label{tab:Sraw}
\footnotesize
\begin{tabular}{lcc}
\toprule
Scenario & GL, raw & TVGL, raw \\
\midrule
S1G & 0.47 (44) & 0.17 (138) \\
S1P & 0.40 (35) & 0.13 (137) \\
S1N & 0.41 (29) & 0.12 (136) \\
S1PF & 0.24 (6) & 0.11 (136) \\
S1GF & 0.29 (8) & 0.15 (138) \\
S1B & 0.30 (13) & 0.06 (136) \\
S2G & 0.70 (66) & 0.30 (130) \\
S2P & 0.66 (72) & 0.22 (129) \\
\bottomrule
\end{tabular}
\end{table}
\begin{table}[t]
\centering
\caption{Prediction of whole withheld maps in all scenarios, at the posterior mean of the expected response, means over ten data sets. IF: independent fields (the fit without a graph, step 1). Joint-0 and Joint: joint EM fits of the static and the switching graph. MCAR-AR1: multivariate CAR model with AR(1) dynamics, run in four scenarios. Div.: divergent predictions.}\label{tab:Spred}
\footnotesize
\begin{tabular}{lccccccc}
\toprule
Scenario & SPDE-AR1 & MCAR-AR1 & IF & Joint-0 & Joint & DBMW-X & DBMW \\
\midrule
S1G & 7.99 & 8.82 & 7.38 & 6.08 & \textbf{5.83} & 6.36 & 5.98 \\
S1P & 4.18 & div. & 4.06 & 3.07 & 2.98 & 3.37 & \textbf{2.96} \\
S1N & 1.897 & -- & 1.863 & 1.729 & 1.722 & 1.742 & \textbf{1.706} \\
S1PF & 4.30 & -- & 4.25 & 4.18 & 4.03 & 3.88 & \textbf{3.77} \\
S1GF & 7.71 & -- & 7.12 & 6.63 & 6.29 & 6.21 & \textbf{6.03} \\
S1B & 1.202 & -- & 1.203 & 1.197 & 1.196 & 1.197 & \textbf{1.195} \\
M1 & 8.03 & div. & 7.09 & 6.41 & \textbf{6.33} & 6.41 & 6.44 \\
M2 & 4.17 & -- & 3.90 & 3.47 & 3.39 & 3.60 & \textbf{3.37} \\
M3 & 4.39 & -- & 4.39 & 3.40 & 3.35 & 3.28 & \textbf{3.13} \\
SCP & 8.13 & div. & 7.55 & \textbf{7.18} & 7.24 & 7.64 & 7.45 \\
S2G & 11.66 & -- & 11.14 & 5.86 & \textbf{5.84} & 6.95 & 6.35 \\
S2P & 3.31 & -- & 3.22 & 2.15 & \textbf{2.13} & 2.61 & 2.27 \\
\bottomrule
\end{tabular}
\end{table}
\begin{table}[t]
\centering
\caption{Uncertainty in the other scenarios, means over ten data sets. ECE: expected calibration error of the inclusion probabilities (ten bins). FDR: realized false discovery rate of the Bayesian rule at $q=0.10$. Correct: share of selected (link, period) pairs with the true sign. False flips: share of single-sign links selected with both signs. Phases found: share of true sign phases, by length in periods, in which the selected graph contains the link with the true sign in at least one period. $\hat\epsilon_{\rm off}$: estimated switching rate, with the realized rate in brackets. Coverage: share of 95\% intervals that contain the realized rate. Bold: the best value in each column, unless all values are equal. For FDR the best value is the closest to the nominal 0.10, and for $\hat\epsilon_{\rm off}$ the closest to the realized rate in relative terms.}\label{tab:Sunc}
\footnotesize
\begin{tabular}{lccccccccc}
\toprule
& \multicolumn{2}{c}{Inclusion} & \multicolumn{2}{c}{Signs} & \multicolumn{3}{c}{Phases found, by length} & \multicolumn{2}{c}{Switching rate} \\
\cmidrule(lr){2-3}\cmidrule(lr){4-5}\cmidrule(lr){6-8}\cmidrule(lr){9-10}
Scenario & ECE & FDR & Correct & False flips & 1--7 & 8--15 & $\geq 16$ & $\hat\epsilon_{\rm off}$ (true) & Coverage \\
\midrule
S1GF & \textbf{0.020} & \textbf{0.11} & 0.995 & \textbf{0.000} & 0.29 & 0.51 & \textbf{0.74} & 0.045 (0.061) & \textbf{1.0} \\
S1B & 0.057 & 0.21 & \textbf{1.000} & \textbf{0.000} & 0.11 & 0.19 & 0.26 & 0.103 (0.099) & 0.9 \\
M1 & 0.042 & 0.32 & \textbf{1.000} & 0.001 & 0.34 & 0.58 & 0.63 & 0.120 (0.099) & 0.9 \\
M2 & 0.038 & 0.15 & \textbf{1.000} & \textbf{0.000} & 0.33 & 0.44 & 0.57 & 0.131 (0.165) & \textbf{1.0} \\
M3 & 0.050 & 0.16 & \textbf{1.000} & \textbf{0.000} & 0.28 & 0.51 & 0.58 & \textbf{0.062 (0.064)} & 0.8 \\
SCP & 0.445 & 1.00 & -- & \textbf{0.000} & -- & -- & -- & 0.015 (--) & -- \\
S2G & 0.078 & 0.06 & \textbf{1.000} & \textbf{0.000} & -- & -- & 0.60 & 0.019 (--) & -- \\
S2P & 0.108 & 0.08 & \textbf{1.000} & \textbf{0.000} & -- & -- & 0.48 & 0.037 (--) & -- \\
S1PFS & 0.025 & 0.16 & 0.998 & 0.001 & \textbf{0.39} & \textbf{0.62} & 0.65 & 0.050 (0.061) & 0.9 \\
\bottomrule
\end{tabular}
\end{table}
\begin{table}[t]
\centering
\caption{Sign phases found in the scenarios with sign changes, by phase length in periods, for links whose truth has both signs (Both) and links with one sign (One), means over ten data sets. A phase is found when the selected graph contains the link with the true sign in at least one period of the phase. Both found: share of links with both signs in the truth that are selected with both signs.}\label{tab:Sphase}
\footnotesize
\begin{tabular}{llcccccccc}
\toprule
& & \multicolumn{2}{c}{1--7} & \multicolumn{2}{c}{8--15} & \multicolumn{2}{c}{$\geq 16$} & \multicolumn{2}{c}{Both found} \\
\cmidrule(lr){3-4}\cmidrule(lr){5-6}\cmidrule(lr){7-8}\cmidrule(lr){9-10}
Scenario & Graph & Both & One & Both & One & Both & One & all & shorter $\geq 8$ \\
\midrule
S1PF & Step 2 & 0.10 & 0.32 & 0.32 & 0.55 & 0.55 & 0.75 & 0.09 & 0.17 \\
 & Step 3 & 0.12 & 0.35 & 0.36 & 0.53 & 0.55 & 0.77 & 0.15 & 0.27 \\
S1GF & Step 2 & 0.16 & 0.41 & 0.44 & 0.57 & 0.61 & 0.83 & 0.15 & 0.30 \\
 & Step 3 & 0.13 & 0.45 & 0.46 & 0.55 & 0.57 & 0.85 & 0.22 & 0.40 \\
S1PFS & Step 2 & 0.26 & 0.51 & 0.52 & 0.72 & 0.52 & 0.74 & 0.18 & 0.27 \\
 & Step 3 & 0.32 & 0.54 & 0.66 & 0.70 & 0.57 & 0.77 & 0.35 & 0.50 \\
\bottomrule
\end{tabular}
\end{table}

\FloatBarrier
\section{Switching-rate study}\label{sec:ratestudy}
Locations, variables, graph and strengths are as in \S\Mref{sec:simdesign}, with Poisson counts and symmetric switching at rate $\epsilon$. Around the baseline $n=200$, $T=52$, $p=20$, one quantity varies at a time: $n\in\{100,200,400,800\}$, $T\in\{26,52,104\}$ and $p\in\{10,20,40\}$, with $\epsilon\in\{0.05,0.1,0.2\}$ and ten data sets per combination. Step~1 runs exactly as for an application. Table~\ref{tab:Srate} gives the results. The estimate tends to fall below the realized rate, most at $\epsilon=0.2$, where short phases are hard to separate from noise. Coverage is 70--100\%, and lowest for the smallest $n$ and the largest $p$ at $\epsilon=0.2$.

\begin{table}[t]
\centering
\caption{Switching-rate study: realized rate, estimate $\hat\epsilon_{\rm off}$ and coverage of the 95\% interval for the realized rate, ten data sets per row. Baseline $n=200$, $T=52$, $p=20$; one quantity varies at a time.}\label{tab:Srate}
\footnotesize
\begin{tabular}{rrrrccc}
\toprule
$\epsilon$ & $n$ & $T$ & $p$ & Realized & $\hat\epsilon_{\rm off}$ & Coverage \\
\midrule
0.05 & 100 & 52 & 20 & 0.051 & 0.053 & 0.9 \\
0.05 & 200 & 26 & 20 & 0.048 & 0.037 & 1.0 \\
0.05 & 200 & 52 & 10 & 0.051 & 0.040 & 1.0 \\
0.05 & 200 & 52 & 20 & 0.045 & 0.041 & 1.0 \\
0.05 & 200 & 52 & 40 & 0.047 & 0.038 & 0.9 \\
0.05 & 200 & 104 & 20 & 0.053 & 0.048 & 1.0 \\
0.05 & 400 & 52 & 20 & 0.052 & 0.044 & 1.0 \\
0.05 & 800 & 52 & 20 & 0.049 & 0.045 & 1.0 \\
0.1 & 100 & 52 & 20 & 0.095 & 0.070 & 0.8 \\
0.1 & 200 & 26 & 20 & 0.092 & 0.079 & 1.0 \\
0.1 & 200 & 52 & 10 & 0.106 & 0.182 & 0.9 \\
0.1 & 200 & 52 & 20 & 0.096 & 0.093 & 1.0 \\
0.1 & 200 & 52 & 40 & 0.097 & 0.086 & 1.0 \\
0.1 & 200 & 104 & 20 & 0.100 & 0.090 & 1.0 \\
0.1 & 400 & 52 & 20 & 0.095 & 0.086 & 1.0 \\
0.1 & 800 & 52 & 20 & 0.099 & 0.092 & 1.0 \\
0.2 & 100 & 52 & 20 & 0.194 & 0.143 & 0.8 \\
0.2 & 200 & 26 & 20 & 0.199 & 0.157 & 0.9 \\
0.2 & 200 & 52 & 10 & 0.200 & 0.178 & 1.0 \\
0.2 & 200 & 52 & 20 & 0.188 & 0.173 & 1.0 \\
0.2 & 200 & 52 & 40 & 0.189 & 0.152 & 0.7 \\
0.2 & 200 & 104 & 20 & 0.199 & 0.171 & 0.8 \\
0.2 & 400 & 52 & 20 & 0.190 & 0.170 & 1.0 \\
0.2 & 800 & 52 & 20 & 0.197 & 0.188 & 1.0 \\
\bottomrule
\end{tabular}
\end{table}

\FloatBarrier
\section{Hold-out designs and a vanishing variable}\label{sec:holdout}
Besides whole maps, two hold-out designs were run in S1G, S1P and S2P. \emph{Regions} splits the locations into ten spatial regions by $k$-means and withholds, for each variable, one region at all times. \emph{Cells} withholds 10\% of the (location, variable, time) entries at random. Table~\ref{tab:Smask} gives the losses. Under a switching truth, DBMW has the lowest loss in all three designs. Under a static truth, Joint-0 has the lowest loss.

\begin{table}[t]
\centering
\caption{Hold-out designs in the simulations: loss at the posterior mean, means over ten data sets. Regions: one spatial region per variable withheld at all periods. Cells: 10\% of the entries at random. Joint-0: joint EM fit of the static graph.}\label{tab:Smask}
\footnotesize
\begin{tabular}{llcccc}
\toprule
Scenario & Hold-out & IF & Joint-0 & DBMW-X & DBMW \\
\midrule
S1G & whole maps & 7.38 & 6.08 & 6.36 & \textbf{5.98} \\
 & regions & 6.48 & 6.23 & 6.26 & \textbf{6.13} \\
 & cells & 3.06 & 3.04 & 3.05 & \textbf{3.03} \\
S1P & whole maps & 4.06 & 3.07 & 3.37 & \textbf{2.96} \\
 & regions & 3.40 & 3.26 & 3.33 & \textbf{3.25} \\
 & cells & 1.731 & 1.723 & 1.729 & \textbf{1.720} \\
S2P & whole maps & 3.22 & \textbf{2.15} & 2.61 & 2.27 \\
 & regions & 2.53 & \textbf{2.34} & 2.49 & 2.44 \\
 & cells & 1.551 & \textbf{1.535} & 1.546 & 1.543 \\
\bottomrule
\end{tabular}
\end{table}

In a further study, one variable of S1P had its rates multiplied by 0.02 in the middle third of the periods, about 91\% zero counts, with no change to its true links. Table~\ref{tab:Svanish} gives the mean inclusion probability of its true links inside and outside this window, and the Brier score of all its links. With the drop in the offset, the Step-2 probabilities are the same inside and outside the window. A vanishing variable therefore produces no spurious switch-off. Without the drop in the offset, the field absorbs an unexplained level shift, and the probabilities fall at all times. The Step-2 graph has the lowest Brier score in both cases.

\begin{table}[!htbp]
\centering
\caption{A vanishing variable in S1P: mean inclusion probability of its true links inside and outside the window of near-zero counts, and Brier score, means over ten data sets. Joint: joint EM fit of the switching graph.}\label{tab:Svanish}
\footnotesize
\begin{tabular}{lcccccc}
\toprule
& \multicolumn{3}{c}{Drop in the offset} & \multicolumn{3}{c}{Drop not in the offset} \\
\cmidrule(lr){2-4}\cmidrule(lr){5-7}
& Inside & Outside & Brier & Inside & Outside & Brier \\
\midrule
Step 2 & 0.39 & 0.39 & 0.168 & 0.13 & 0.16 & 0.225 \\
Step 3 & 0.37 & 0.35 & 0.205 & 0.05 & 0.13 & 0.255 \\
Joint & 0.15 & 0.21 & 0.247 & 0.00 & 0.05 & 0.259 \\
\bottomrule
\end{tabular}
\end{table}

\FloatBarrier
\section{Competitors}\label{sec:competitors}
The competitor SPDE-AR1 gives each variable one Mat\'ern SPDE field with AR(1) evolution over the periods \citep{cameletti2013}, fitted by R-INLA \citep{rue2009, lindgren2011} with the Gaussian approximation strategy and empirical Bayes estimates of the hyperparameters. The priors are penalized-complexity priors with $\Pr(\text{range}<0.2)=0.5$ and $\Pr(\sigma>2)=0.01$ on the coordinates scaled to the unit square, and a penalized-complexity prior on the AR(1) correlation with $\Pr(\rho>0.5)=0.7$. The default mesh has maximal edge lengths 0.1 inside and 0.3 outside the domain, with cutoff 0.045. The responses, offsets and hold-out entries are those of our fits. The withheld entries are scored at the posterior mean of the expected response.

In the applications the default mesh has fewer nodes than locations: 235 nodes for the 288 air-quality sites and 299 for the 411 districts. Finer meshes lower the error on whole air-quality maps from 0.334 to 0.272 with 762 nodes and to 0.232 with 1899 nodes, at 26 and 231 minutes per fit. In fit A of the surveillance data, a mesh with 984 nodes lowers the deviance from 7.23 to 6.70, between DBMW (6.33) and DBMW-X (7.28), at 314 minutes per fit. All other comparisons use the default mesh.

In the three withheld-region data sets of the air-quality application, SPDE-AR1 fits the \NOt\ field with an AR(1) correlation of 0.9998 and a field standard deviation of 8.0--8.9. The intercept is then weakly identified. In data sets 1 and 3, the predictions in the withheld \NOt\ region are biased by about $-8$ standardized units, with squared errors of 81 and 70. In data set 2 the \NOt\ error is 1.7. The mean squared error over all withheld entries is 18.3, 2.9 and 25.9 in the three data sets.

We therefore also fit SPDE-AR1 with an added time-constant spatial field for each variable, with the same penalized-complexity priors as the AR(1) field. For \NOt, the AR(1) correlation then falls to 0.97--0.99 with a field standard deviation below 1. The site field takes a range of about 0.05 and a standard deviation of 4.4--22. The \NOt\ errors fall to 7.6 and 1.3 in data sets 1 and 2, but they stay at 64 in data set 3. On a mesh of 235 nodes, a Mat\'ern field can represent levels that differ between nearby monitors only with a very short range and a large variance, and its predictions in a withheld region are then unstable. Table~\Mref{tab:app} reports both specifications. The site field lowers the error with withheld regions in the air-quality data and in fit A, raises it with withheld regions in fit B, and changes the whole-map errors by at most 5\%.

The multivariate CAR model with AR(1) dynamics \citep{lee2018carbayesst} is \texttt{MVST.CARar} of CARBayesST, with an unstructured $p\times p$ covariance between the variables, a Leroux CAR on the 10-nearest-neighbour graph and AR(1) evolution, fitted by MCMC with 10{,}000 burn-in and 30{,}000 iterations. It was run in four scenarios. On whole withheld count maps it diverges (Table~\ref{tab:Spred}). In the surveillance application it diverges in fit A and reaches a deviance of 550 in fit B. On whole air-quality maps its error is 0.303.

\FloatBarrier
\section{Applications}\label{sec:apps}
\begin{figure}[!htbp]
\centering
\includegraphics[width=\linewidth]{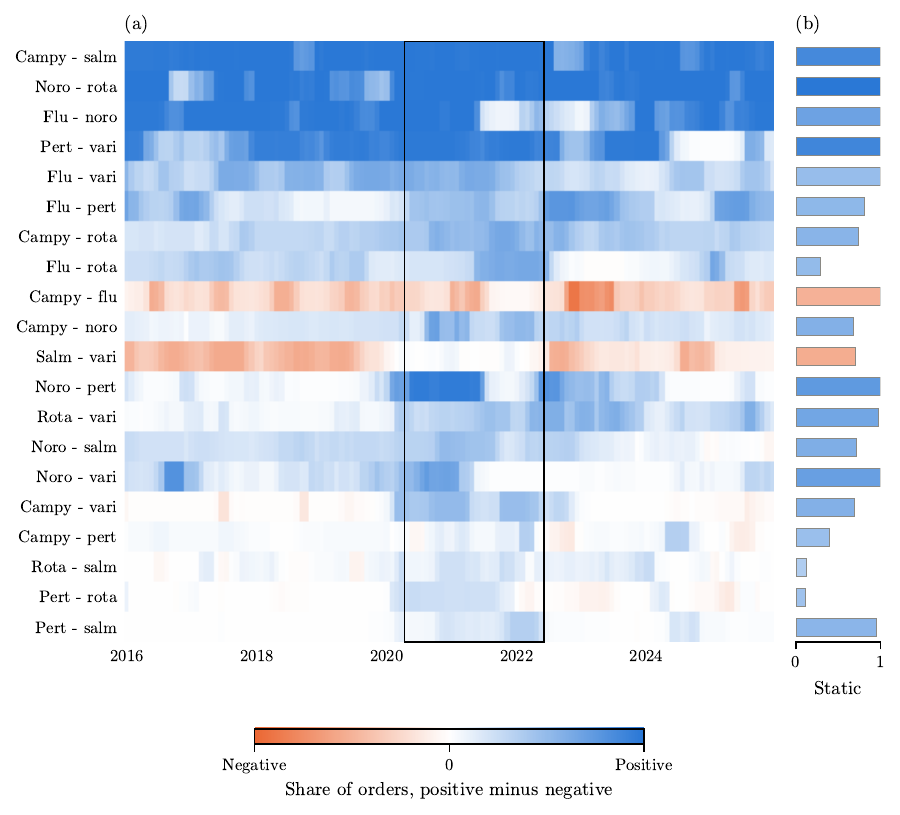}
\caption{Infectious disease surveillance, fit A (population offset). (a) Share of the 120 variable orders in which the Step-2 graph selects each link with a positive association, minus the share with a negative association, by four-week period. The frame marks the pandemic period. (b) Share of the orders in which the static graph of DBMW-0 selects each link. Abbreviations as in Figure~\Mref{fig:survapp}.}\label{fig:survA}
\end{figure}

Table~\ref{tab:Sappdata} summarizes the two data sets. Table~\ref{tab:Sapp} reports prediction by all estimators. Figure~\ref{fig:survA} shows the signed graph of fit A of the surveillance data.

\begin{table}[!htbp]
\centering
\caption{The two data sets.}\label{tab:Sappdata}
\footnotesize
\begin{tabular}{p{3.0cm}p{5.3cm}p{5.3cm}}
\toprule
& Air quality & Infectious disease surveillance \\
\midrule
Source & EPA Air Quality System, daily summaries & SurvStat@RKI 2.0, notified cases \\
Sites & 288 monitoring sites, contiguous U.S. & 411 districts, Germany \\
Variables ($p$) & temperature, \NOt, CO, \PMt, ozone (5) & influenza, norovirus, rotavirus, pertussis, varicella, campylobacteriosis, salmonellosis (7) \\
Periods ($T$) & 52 weeks of 2023 & 130 four-week periods, 2016--2025 \\
Response, family & weekly mean, log for pollutants, standardized, Gaussian & case count, Poisson \\
Offset & none & A: log(population $\times$ weeks), B: A plus national log rate \\
Observed entries & 79, 98, 34, 61, 98\% by variable & all, with 3--44\% zeros by disease \\
Candidate pairs & all 10 & all 21 \\
Orders & all 120 & the data order and 119 random orders \\
Outer iterations & 20 & 15 \\
\bottomrule
\end{tabular}
\end{table}

In the air-quality data, DBMW and DBMW-X predict best on whole maps. Over the 120 orders, the median error of DBMW-X is lower than that of DBMW. With withheld regions and scattered cells, the estimators with a graph differ by less than 3\%. In fit A of the surveillance data, the static graph predicts whole maps best, and the joint EM fit predicts withheld regions best, with DBMW within 1\%. On withheld regions, DBMW beats the static graph in all three data sets of fit A. In fit B, the fit without a graph predicts whole maps as well as any graph, because the national curves carry the timing. With withheld regions, every graph lowers the deviance by about 6\% relative to the fit without a graph.

\begin{table}[t]
\centering
\caption{Prediction in the applications by all estimators, at the posterior mean. IF: independent fields. DBMW-0: static graph. Joint-0 and Joint: joint EM fits of the static and the switching graph. Cells: 10\% of the entries at random, means over three data sets. Other rows as in Table~\ref{tab:app}.}\label{tab:Sapp}
\footnotesize
\begin{tabular}{llcccccc}
\toprule
Data & Hold-out & IF & DBMW-0 & Joint-0 & Joint & DBMW-X & DBMW \\
\midrule
Air quality & whole maps & 0.213 & 0.200 & 0.200 & 0.197 & 0.188 & \textbf{0.187} \\
 & whole maps, orders & 0.213 & 0.199 & -- & -- & \textbf{0.190} & 0.196 \\
 & regions & 0.861 & 0.718 & 0.717 & \textbf{0.704} & 0.720 & 0.715 \\
 & cells & 0.081 & 0.078 & 0.078 & 0.078 & \textbf{0.077} & 0.078 \\
\addlinespace
Surveillance, fit A & whole maps & 8.13 & 5.87 & \textbf{5.81} & 6.37 & 7.28 & 6.33 \\
 & regions & 12.06 & 15.39 & 13.10 & \textbf{11.60} & 12.31 & 11.70 \\
 & cells & 3.12 & 3.11 & \textbf{3.07} & \textbf{3.07} & 3.11 & \textbf{3.07} \\
\addlinespace
Surveillance, fit B & whole maps & \textbf{9.57} & 9.67 & 9.60 & 9.61 & 9.68 & 9.60 \\
 & regions & 5.84 & 5.51 & 5.45 & \textbf{5.44} & 5.53 & 5.48 \\
 & cells & 3.05 & 3.03 & \textbf{3.02} & \textbf{3.02} & 3.03 & \textbf{3.02} \\
\bottomrule
\end{tabular}
\end{table}

In fit A, the fields carry the national seasons, and links switch more often than in fit B, with $\hat\epsilon_{\rm off}=0.117$ against 0.042 per period. Two links have a negative association: campylobacteriosis with influenza, in up to 90\% of the orders, and salmonellosis with varicella, in up to 56\%. The enteric bacteria peak in summer, influenza in winter and varicella in spring, so the negative links reflect opposite national seasons rather than district-level dependence. In fit B both links lose their negative association.

\end{document}